\documentclass[10pt,journal,letterpaper,compsoc]{IEEEtran}
\usepackage{amssymb}
\usepackage{times}
\usepackage{graphicx}
\usepackage{amsthm}
\usepackage{array}
\usepackage{float}
\usepackage{multirow}
\usepackage{subfig}
\usepackage{amsmath}
\usepackage{mathabx}
\usepackage{algorithm}
\usepackage{algpseudocode}
\usepackage{url}

\begin{document}
	
	%
	\title{A Survey on Spark Ecosystem for Big Data Processing}

\author{Shanjiang Tang, Bingsheng He, Ce Yu, Yusen Li, Kun Li
	\IEEEcompsocitemizethanks{\IEEEcompsocthanksitem  S.J. Tang, C. Yu, K. Li are with the College of Intelligence and Computing, Tianjin University,
		Tianjin 300072, China.
		\protect\\
		E-mail: \{tashj, yuce, kunli\}@tju.edu.cn.
		\IEEEcompsocthanksitem B.S. He is with the School of Computing, National University of Singapore.
		\protect\\
		E-mail: hebs@comp.nus.edu.sg
		\IEEEcompsocthanksitem Yusen Li is with the School of Computing, Nankai University, Tianjin 300071, China.
		\protect\\
		E-mail:  liyusen@nbjl.nankai.edu.cn.}
	\thanks{}}

%

\markboth{}%
{Shell \MakeLowercase{\textit{et al.}}: Bare Demo of IEEEtran.cls for Computer Society Journals}

\IEEEcompsoctitleabstractindextext{%
	\begin{abstract}
	With the explosive increase of big data in industry and academic fields, it is necessary to apply large-scale data processing systems to analysis Big Data. Arguably, Spark is state of the art in large-scale data computing systems nowadays, due to its good properties including generality, fault tolerance, high performance of in-memory data processing, and scalability. Spark adopts a flexible Resident Distributed Dataset (RDD) programming model with a set of provided transformation and action operators whose operating functions can be customized by users according to their applications. It is originally positioned as a \emph{fast} and \emph{general} data processing system. A large body of research efforts have been made to make it more efficient (faster) and general by considering various circumstances since its introduction.
In this survey, we aim to have a thorough review of various kinds of optimization techniques on the generality and performance improvement of Spark. We introduce Spark programming model and computing system, discuss the pros and cons of Spark, and have an investigation and classification of various solving techniques in the literature. Moreover, we also introduce various data management and processing systems, machine learning algorithms and applications supported by Spark. Finally, we make a discussion on the open issues and challenges for large-scale in-memory data processing with Spark.

\end{abstract}

}
\maketitle
\IEEEdisplaynotcompsoctitleabstractindextext

\IEEEpeerreviewmaketitle

\section{Introduction}

%

In the current era of `big data', the data is collected at unprecedented scale in many application domains, including e-commerce~\cite{Lam:2012:MMP:2367502.2367520}, social network~\cite{Pujol:2010:LES:1851182.1851227}, and computational biology~\cite{Tang:2012:EEP:2197076.2197192}. Given the characteristics of the unprecedented
amount of data, the speed of data production, and the multiple of the structure of data, large-scale data processing is essential to analyzing and mining such big data timely.
A number of large-scale data processing frameworks have thereby been developed, such as MapReduce~\cite{Dean}, Storm~\cite{Storm}, Flink~\cite{flink},  Dryad~\cite{Isard:2007:DDD:1272996.1273005},  Caffe~\cite{jia2014caffe}, Tensorflow~\cite{abadi2016tensorflow}. Specifically, MapReduce is a batch processing framework, while Storm and Flink are both streaming processing systems. Dryad is a graph processing framework for graph applications. Caffe and Tensorflow are deep learning frameworks used for model training and inference in computer vision, speech recognition and natural language processing.

However, all of the aforementioned frameworks are not \emph{general} computing systems since each of them can only work for a certain data computation. In comparison, Spark~\cite{Spark} is a \emph{general} and \emph{fast} large-scale data processing system widely used in both industry and academia with many merits.
For example, Spark is much faster than MapReduce in performance, benefiting from its in-memory data processing. Moreover, as a general system, it can support batch, interactive, iterative, and streaming computations in the same runtime, which is useful for complex applications that have different computation modes.

Despite its popularity, there are still many limitations for Spark. For example, it requires considerable amount of learning and programming efforts under its RDD programming model. It does not support new emerging heterogenous computing platforms such as GPU and FPGA by default. Being as a general computing system, it still does not support certain types of applications such as deep learning-based applications~\cite{CaffeOnSpark}.

To make Spark more \emph{general} and \emph{fast}, there have been a lot of work made to address the limitations of Spark~\cite{Lu,Davidson,GraphX,Tachyon} mentioned above, and it remains an active research area. A number of efforts have been made on performance optimization for Spark framework. There have been proposals for more complex scheduling strategies~\cite{Sparrow,Shivaram} and efficient memory I/O support (e.g., RDMA support) to improve the performance of Spark. There have also been a number of studies to extend Spark for more sophisticated algorithms and applications (e.g., deep learning algorithm, genomes, and Astronomy). To improve the ease of use, several high-level declarative~\cite{Shark,Apache_Hive,Spark_SQL} and procedural languages~\cite{Spork,Python_Spark} have also been proposed and supported by Spark.

\begin{figure}[h]
	\begin{center}
		\includegraphics
		[width=0.5\textwidth]{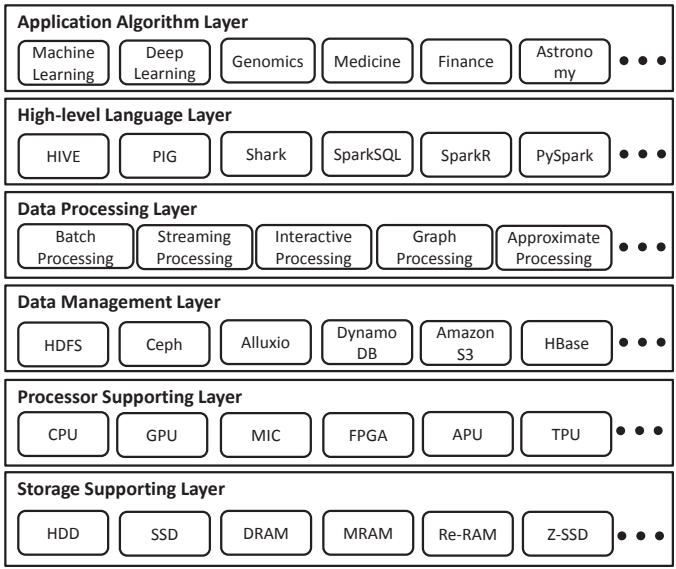}
		\vspace*{-0.1in}
		\caption{\small Overview of Spark ecosystem from the bottom up. We classify it into six layers for improved generality and performance efficiency. }
		\label{Spark_Support}
		\vspace{-0.1in}
	\end{center}
\end{figure}

Still, with the emergence of new hardware, software and application demands, it brings new opportunities as well as challenges to extend Spark for improved generality and performance efficiency. In this survey, for the sake of better understanding these potential demands and opportunities systematically, we classify the study of Spark ecosystem into six support layers as illustrated in Figure~\ref{Spark_Support}, namely, Storage Supporting Layer, Processor Supporting Layer, Data Management Layer, Data Processing Layer, High-level Language Layer and Application Algorithm Layer. The aim of this paper is two-fold. We first seek to have an investigation of the latest studies on Spark ecosystem. We review related work on Spark and classify them according to their optimization strategies in order to serve as a guidebook for users on the problems and addressing techniques in data processing with Spark. It summarizes existing techniques systematically as a dictionary for expert researchers to look up. Second, we show and discuss the development trend, new demands and challenges at each support layer of Spark ecosystem as illustrated in Figure~\ref{Spark_Support}. It provides researchers with insights and potential study directions on Spark.


The rest part of this survey is structured as follows. Section~\ref{Spark_Overview} introduces Spark system, including its programming model, runtime computing engine, pros and cons, and various optimization techniques. Section~\ref{Storage_as_a_Support} describes new caching devices for Spark in-memory computation. Section~\ref{Processor_as_a_Support} discusses the extensions of Spark for performance improvement by using new accelerators. Section~\ref{Data_Management_as_a_Support} presents distributed data management, followed by processing systems supported by Spark in Section~\ref{Processing_as_a_Support}. Section~\ref{Language_as_a_Support} shows the languages that are supported by Spark. Section~\ref{Application_as_a_Support} reviews the Spark-based machine learning libraries and systems, Spark-based deep learning systems, and the major applications that the Spark system is applied to. Section~\ref{Challenges_and_Open_Issues} makes some open discussion on the challenging issues. Finally, we conclude this survey in Section~\ref{Conclusion}.

%
%

\section{Core Techniques of Spark}
\label{Spark_Overview}
This section first describes the RDD programming model, followed by the overall architecture of Spark framework. Next it shows the pros and cons of Spark, and various optimization techniques for Spark.

\subsection{Programming Model}

Spark is based on Resilient Distributed Dataset (RDD)~\cite{RDD} abstraction model, which is an immutable collection of records partitioned across a number of computers. Each RDD is generated from data in external robust storage systems such as HDFS, or other RDDs through coarse-grained \emph{transformations} including \emph{map}, \emph{filter} and \emph{groupByKey} that use identical processing to numerous data records. To provide fault tolerance, each RDD's transformation information is logged to construct a lineage dataset. When a data partition of a RDD is lost due to the node failure, the RDD can recompute that partition with the full information on how it was generated from other RDDs. It is worthy mentioning that the transformation is a \emph{lazy} operation that only defines a new RDD instead of calculating it immediately. In order to launch the computation of RDD, Spark offers another group of \emph{action} operations such as \emph{count}, \emph{collect}, \emph{save} and \emph{reduce}, which either return a value to an application program or export the RDD's data to an external storage system. Moreover, for the data of a RDD, they can be persisted either in memory or in disk, controlled by users.

\subsection{Spark Architecture}

\begin{figure}[h]
	\begin{center}
		\includegraphics
		[width=0.5\textwidth]{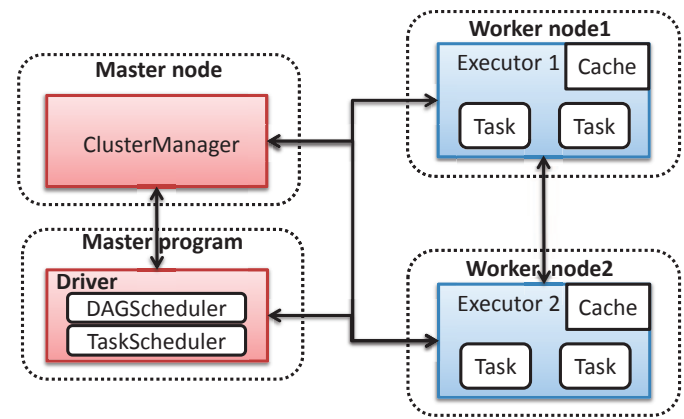}
		\vspace*{-0.05in}
		\caption{\small Architecture overview of Spark.}
		\label{Spark_Architecture}
		\vspace{-0.1in}
	\end{center}
\end{figure}

Figure~\ref{Spark_Architecture} overviews the architecture of Spark on a cluster. For each Spark application, it spawns one master process called \emph{driver}, which is responsible for task scheduling. It follows a hierarchical scheduling process with jobs, stages and tasks, where \emph{stages} refer to as smaller sets of tasks divided from interdependent jobs, which resemble map and reduce phases of a MapReduce job. There are two schedulers inside it, namely, \emph{DAGScheduler} and \emph{TaskScheduler}. The DAGScheduler computes a DAG of stages for a job and keeps track of the materialized RDDs as well as stage outputs, whereas TaskScheduler is a low-level scheduler that is responsible for getting and submitting tasks from each stage to the cluster for execution.

Spark provides users with three different cluster modes (i.e., Mesos~\cite{Mesos}, YARN~\cite{YARN}, and standalone mode) to run their Spark applications by allowing driver process to connect to one of existing popular cluster managers including Mesos, YARN and its own independent cluster manager. In each worker node, there is a slave process called \emph{executor} created for each application, which is responsible for running the tasks and caching the data in memory or disk.

\subsection{Pros and Cons of Spark}
\label{Spark_Pros_Cons}
MapReduce was a powerful large-scale data processing system widely used for many data-intensive applications. In this section, we take MapReduce as a baseline to discuss the pros and cons of Spark.

\subsubsection{Strength}

\emph{Easy to use}. Spark provides users with more than 80 high-level simple operators (e.g., \emph{map}, \emph{reduce}, \emph{reduceByKey}, \emph{filter}) that allow users to write parallel applications at the application level with no need to consider the underlying complex parallel computing problems like data partitioning, task scheduling and load balancing. Moreover, Spark allows users to write their user-defined functions with different programming languages like Java, Scala, Python by offering corresponding APIs.

\emph{Faster than MapReduce}. Due to its in-memory computing, Spark has shown to be $10\times \sim 100\times$ faster than MapReduce in  batch processing~\cite{Spark_homepage}.

\emph{General computation support}. First, from the aspect of processing mode, Spark is an integrated system that supports batch, interactive, iterative, and streaming processing. Second, Spark has an advanced DAG execution engine for complex DAG applications, and a stack of high-level APIs and tools including Shark~\cite{Shark}, Spark SQL~\cite{Spark_SQL}, MLlib and Graphx~\cite{GraphX} for a wide range of applications.

\emph{Flexible running support}. Spark can run in a standalone mode or share the cluster with other computing systems like MapReduce by running on YARN or Mesos. It also provides APIs for users to deploy and run on the cloud (e.g., Amazon EC2). Moreover, it can support the access of various data sources including HDFS, Tachyon~\cite{Tachyon}, HBase, Cassandra~\cite{Cassandra}, and Amazon S3~\cite{Amazon_S3}.

\subsubsection{Weakness}
%
Albeit many benefits, there are still some weakness for Spark, compared with MapReduce as follows:

\emph{Heavy consumption of storage resources}. As an in-memory data processing framework, Spark is superior to MapReduce in performance, achieved by reducing the redundant computations at the expense of storage resources, especially memory resource. Similar to existing popular in-memory caching systems like Memcached~\cite{Nishtala, Zhang} and Redis~\cite{Redis}, it stores RDD data in memory and keeps it there for data sharing across different computation stages. More memory resources are needed when there are a large volume of RDD data to be cached in computation.

\emph{Poor security}. Currently, Spark supports authentication through a shared secret~\cite{Spark_Security}. In comparison, Hadoop has more security considerations and solutions, including Knox~\cite{Knox}, Sentry~\cite{Sentry}, Ranger~\cite{Apache_Ranger}, etc. For example, Knox provides the secure REST API gateway for Hadoop with authorization and authentication. In contrast, Sentry and Ranger offer access control and authorization over Hadoop data and metadata.


\emph{Learning Curve}. Although Spark is faster and more general than MapReduce, the programming model of Spark is much more complex than MapReduce. It requires users to take time to learn the model and be familiar with provided APIs before they can program their applications with Spark.

\subsubsection{Comparison}

\begin{table}[htbp]\scriptsize
\begin{center}
\hspace*{-0.15in}
\begin{tabular}{|c|c|c|}
\hline
\textbf{Metrics} & \textbf{Spark} & \textbf{MapReduce} \\
\hline\hline
Usability & Easy-to-use & Easy-to-use \\
\hline
Performance & High Efficiency & Low Efficiency \\
\hline
Generality & Yes & No \\
\hline
Flexibility & Yes & Yes \\
\hline
Scalability & Yes & Yes \\
\hline
Fault Tolerance & Yes & Yes \\
\hline
Memory Consumption & Heavy & Heavy \\
\hline
Security & Poor & Strong \\
\hline
Learning & hard-to-learn & easy-to-learn \\
\hline
\end{tabular}
\caption{\small The comparison of Spark and MapReduce.}
\label{Spark_MapReduce_Pros_Cons}
\end{center}
\vspace*{-0.1in}
\end{table}

For the sake of better understanding Spark's characteristics, we make a comparison of Spark and MapReduce in Table~\ref{Spark_MapReduce_Pros_Cons} with respect to different metrics. First, both frameworks have a good usability, flexibility, scalability, and fault tolerance properties. All of complex details of distributed computation are encapsulated and well considered by frameworks and are transparent to users. Second, Spark is superior to MapReduce in performance and generality, attributing to Spark's in-memory computation and RDD programming model. Reversely, MapReduce has a stronger security and easy-to-learn property than Spark. Compared to Spark, the programming model of MapReduce is more simple and mature. Finally, both frameworks have the problem of high memory consumption, due to the heavy memory usage of JVMs. Particularly, for Spark, its in-memory RDD caching consumes a large amount of memory resources.

\subsection{Spark System Optimization}
\label{Spark_Improvement}

Performance is the most important concern for Spark system. Many optimizations are studied on top of Spark in order to accelerate the speed of data handling. We mainly describe the major optimizations proposed on the Spark system in this section.

\subsubsection{Scheduler Optimization}
The current Spark has a centralized scheduler which allocates the available resources to the pending tasks according to some policies (e.g., FIFO or Fair). The design of these scheduling policies can not satisfy the requirements of current data analytics. In this section, we describe different kinds of schedulers that are especially optimized for large-scale distributed scheduling, approximate query processing, transient resource allocation and Geo-distributed setting, respectively.

\emph{Decentralized Task Scheduling}. Nowadays, more and more Big Data analytics frameworks are moving towards larger degrees of parallelism and shorter task durations in order to provide low latency. With the increase of tasks, the throughput and availability of current centralized scheduler can not offer low-latency requirement and high availability. A decentralized design without centralized state is needed to provide attractive scalability and availability. Sparrow~\cite{Sparrow} is the-state-of-art distributed scheduler on top of Spark. It provides the power of two choices load balancing technique for Spark task scheduling. The power probes two random servers and places tasks on the server with less load. Sparrow makes the power of two choices technique effective in parallel jobs running on a cluster with the help of three techniques, namely, Batch Sampling, Late Binding and Policies and Constraints. Batch Sampling reduces the time of tasks response which is decided by the finishing time of the last task by placing tasks of one job in a batch way instead of sampling for each task individually. For the power of two choices, the length of server queue is a poor norm of latency time and the parallel sampling may cause competition. Late binding prevents two issues happening by delaying allocation of tasks to worker nodes before workers get ready to execute these tasks. Sparrow also enforces global policies using multiple queues on worker machines and supports placement constraints of each job and task.

\emph{Data-aware Task Scheduling}. For machine learning algorithms and sampling-based approximate query processing systems, the results can be computed using any subset of the data without compromising application correctness. Currently, schedulers require applications to statically choose a subset of the data that the scheduler runs the task which aviods the scheduler leveraging the combinatorial choices of the dataset at runtime. The data-aware scheduling called KMN~\cite{Shivaram} is proposed in Spark to take advantage of the available choices. KMN applies the ``late binding" technique which can dynamically select the subset of input data on the basis of the current cluster's state. It significantly increases the data locality even when the utilization of the cluster is high. KMN also optimizes for the intermediate stages which have no choice in picking their input because they need all the outputs produced by the upstream tasks. KMN launches a few additional jobs in the previous stage and pick choices that best avoid congested links.

\emph{Transient Task Scheduling}. For cloud servers, due to various reasons, the utilization tends to be low and raising the utilization rate is facing huge competitive pressure. One addressing solution is to run insensitive batch job workloads secondary background tasks if there are under-utilized resources and evicted them when servers's primary tasks requires more resources (i.e., \emph{transit resources}). Due to excessive cost of cascading re-computations, Spark works badly in this case. TR-Spark (Transient Resource Spark)~\cite{Yan:2016:TTC:2987550.2987576} is proposed to resolve this problem. It is a new framework for large-scale data analytic on transient resources which follows two rules: data scale reduction-aware scheduling and lineage-aware checkpointing. TR-Spark is implemented by modifying Spark's Task Scheduler and Shuffle Manager, and adding two new modules Checkpointing Scheduler and Checkpoint Manager.

\emph{Scheduling in a Geo-distributed Environment}. Geo-distributed data centers are deployed globally to offer their users access to services with low-latency. In Geo-distributed setting, the bandwidth of WAN links is relatively low and heterogeneous compared with the intra-DC networks. The query response time over the current intra-DC analytics frameworks becomes extreme high in Geo-distributed setting. Iridium~\cite{pu2015low} is a system designed for Geo-distributed data analytics on top of Spark. It reduces the query response time by leveraging WAN bandwidth-aware data and task placement approaches. By observing that network bottlenecks mainly occur in the network connecting the data centers rather than in the up/down links of VMs as assumed by Iridium,  Hu \emph{et al.}~\cite{8107572} designed and implemented a new task scheduling algorithm called Flutter on top of Spark. which reduces both the completion time and network costs by formulating the optimization issue as a lexicographical min-max integer linear programming (ILP) problem.

\subsubsection{Memory Optimization}

Efficient memory usage is important for the current in-memory computing systems. Many of these data processing frameworks are designed by garbage-collected languages like C\#, Go, Java or Scala. Unfortunately, these garbage-collected languages are known to cause performance overhead due to GC-induced pause. To address the problem, current studies either improvement the GC performance of these garbage-collected language or leverage application semantics to manage memory explicitly and annihilate the GC overhead of these garbage-collected languages~\cite{spark2,tungsten,maas2016taurus,maas2015trash}. In this section, we introduce these optimizations from these two aspects.

Spark run multiple work processes on different nodes and the Gargabe Collection (GC) is performed independently in each node at run. Works communicate data between different nodes (e.g, shuffle operation). In this case, no node can continue until all data are received from all the other nodes. GC pauses can lead to unacceptable long waiting time for latency-critical applications without the central coordination. If even a single node is stuck in GC, then all the other nodes need wait. In order to coordinate the GC from the central view, Holistic Runtime System~\cite{maas2015trash,maas2016taurus} is proposed to collectively manages runtime GC across multiple nodes. Instead of making decisions about GC independently, such Holistic GC system allows the runtime to make globally coordinated consensus decision through three approaches. First, it let applications choose the most suitable GC policy to match the requirement of different applications (e.g., throughput vs pause times). Second, Holistic system performs GC by considering the application-level optimizations. Third, the GC system is dynamically reconfigured at runtime to adapt to system changes.

Instead of replying the memory management of such managed languages. Spark also tries to manage the memory by itself to leverage the application semantic and eliminate the GC overhead of these garbaged-collected languages.
Tungsten~\cite{tungsten} improves the memory and CPU efficiency of spark applications to make the performance of Spark reach the limits of modern hardware. This work consists of three proposes. First, it leverages the off-heap memory, a feature provided by JAVA to allocate/deallocate memory like c and c++, to manage memory by itself which can take advantage of the application semantics and annihilate the overhead of JVM and GC. Second, it proposes cache-obvious algorithms and data structures to develop memory hierarchical structure. Third, it uses the code generation to avoid the overhead the expression evaluation on JVM (e.g., too many virtual functions calls, extensive memory access and can not take advantage modern CPU features such as SIMD, pipeline and prefetching). Recently, Spark further optimizes its performance by integrating the techniques proposed in Modern parallel database area~\cite{neumann2011efficiently}. Spark 2.0 leverages whole process code generation and vectorization to further ameliorate the code generation at runtime~\cite{spark2}.

\subsubsection{I/O Optimization}

For large-scale data-intensive computation in Spark, the massive data loading (or writing) from (or to) disk, and transmission between tasks at different machines are often unavoidable. A number of approaches are thereby proposed to alleviate it by having a new storage manner, using data compression, or importing new hardware.

\emph{Data Compression and Sharing}. One limitation for Spark is that it can only support the in-memory data sharing for tasks within an application, whereas not for tasks from multiple applications. To overcome this limitation, Tachyon~\cite{Tachyon, Tachyon1} is proposed as a distributed in-memory file system that achieves reliable data sharing at memory speedup for tasks from different processes. The Spark applications can then share their data with each other by writing (or reading) their data to (or from) Tachyon at memory speedup, which is faster than disk-based HDFS file system.  Moreover, to enable more data stored in memory for efficient computation, Agarwal \emph{et al.}~\cite{Succinct} proposed and implemented a distributed data store system called Succinct in Tachyon that compresses the input data and queries can be executed directly on the compressed representation of input data, avoiding decompression.

%

\emph{Data Shuffling}. Besides the performance degradation from the disk I/O, the network I/O may also be a serious bottleneck for many Spark applications. Particularly, \emph{shuffle}, a many-to-many data transfer for tasks across machines, is an important consumer of network bandwidth for Spark. Zhang \emph{et al.}~\cite{Zhang:2018:ROS:3190508.3190534} observed that the bottleneck for shuffle phase is due to large disk I/O operations. To address it, a framework called Riffle is proposed to improve I/O efficiency by merging fragmented intermediate shuffle files into larger block files and converts small and random disk I/O operations into large and sequential ones. Davidson \emph{et al.}~\cite{Davidson} proposed two approaches to optimize the performance in data shuffling. One is to apply the Columnar compression technique to Spark's shuffle phase in view of its success in a column-oriented DBMS called C-Store~\cite{C_store}, so as to offload some burden from the network and disk to CPU. Moreover, they observe that Spark generates a huge number of small-size shuffle files on both the map and reduce phase, which introduces a heavy burden on operating system in file management. A shuffle file consolidation approach is thereby proposed to reduce the number of shuffle files on each machine.

Moreover, prefetching is an effective technique to hide shuffling cost by overlapping data transfers and the shuffling phase. Current state-of-the-art solutions take
simple mechanisms to determine where and how much data to acquire from, resulting in the performance of sub-optimal and the excessive use of supplemental memory. To address it, Bogdan \emph{et al.}~\cite{Nicolae:2017:LAI:3101627.3101641} proposed an original adaptive shuffle data transfer strategy by dynamically adapting the prefetching to the calculation. It is achieved by taking into account load balancing for request extraction using executor-level coordination, prioritization according to locality and responsiveness, static circular allocation of initial requests, elastic adjustment of in-flight restrictions, shuffle block aggregation and dispersal using in-flight increment.

There are also some work focusing on optimizing shuffling under a certain circumstance. Kim \emph{et al.}~\cite{DBLP:journals/corr/abs-1708-05746} considered the I/O optimization for Spark under large memory servers. It can achieve better data shuffling and intermediate storage by replacing the existing TCP/IP-based shuffle with a large shared memory approach. The communication cost of map and reduce tasks can be reduced significantly through referencing to the global shared memory compared with data transferring over the network.
Liu \emph{et al.}~\cite{7980000} studied the data shuffling in a wide-area network, where data transfers occur between geographically distributed datacenters. It designed and implemented a data aggregation spark-based system by strategically and proactively aggregate the output data of map tasks to a subset of worker datacenters, which replaces the original passive fetch mechanisms used in Spark across datacenters. It can avoid repetitive data transfers and hence improves the utilization of inter-datacenter links.

\emph{RDMA-based Data Transfer}. Lu \emph{et al.}~\cite{Lu} accelerated the network communication of Spark in big data processing using Remote Direct Memory Access (RDMA) technique. They proposed a RDMA-based data shuffle engine for Spark over InfiniBand. With RDMA, the latency of network message communication is dramatically reduced, which improves the performance of Spark significantly.

\subsubsection{Provence Support}
Data-intensive scalable computing (DISC) systems such as Hadoop and Spark, expose a programming model for authoring data processing logic, which is converted to a Directed Acyclic Graph (DAG) of parallel computing~\cite{Interlandi2015Titian}. Debugging data processing logic in DISC systems is difficult and time consuming. A library, \emph{Titian}~\cite{Interlandi2015Titian}, provides data provenance support at the velocity of interactive based on Apache Spark. The contributions of Titian are summarized as follow: A data lineage capture and query support system while minimally impacting Spark job performance. Interactive data provenance query support the expansion of a conversant programming model Spark RDD with less overhead. Titian extends the native Spark RDD interface with tracing capabilities and returns a LineageRDD, traveling by dataflow transformations at stage boundaries. The user is able to retrospect to the intermediate data of the program execution from the given RDD, then leverage local RDD transformations to reprocess the referenced data.

Currently, researchers use cloud computing platforms to analyse Big Data in parallel, but debugging massive parallel computations is time consuming and infeasible for users. To meet the low overhead, scalability and fine-grained demands of big data processing in Apache Spark, a group of interactive and real-time debugging primitives were developed. BIGDEBUG~\cite{Gulzar2016BigDebug} provides simulated breakpoints and guarded watchpoints with the trifling influence of performance, which indicates less than $24\%$ overhead for record-level tracing, $19\%$ overhead for crash monitoring, and $9\%$ overhead for watchpoint on average. BIGDEBUG supports a real-time rapid repair and recovery to prevent re-running the job from the beginning. Besides, BIGDEBUG offers the provenance of the culprit and fine-grained tracking of records in distributed pipes to track intermediate results back and forth.

An improved version of the original Titian system is designed to reduce the lineage query time~\cite{Interlandi2017Adding}. The two key features of Titian are crash culprit determination and automated fault localization. The culprit information is packaged and dispatch to users with other run-time records. The delta debugging technique diagnose whether mistakes in code and data. To promote the performance of lineage queries, they extend Spark with an available way to retrieve lineage records more pragmatically. For large-scale data, small tracing queries generate remarkable overhead from jobs that make little contribution to the result. Therefore, a new custom Spark scheduler, called Hyperdrive, is proposed, which utilizes partition statistics to exclude the situation. Moreover, Hyperdrive decouples task operations from partitions and  dispenses multiple partitions to one task.

\section{Storage Supporting Layer}
\label{Storage_as_a_Support}

\begin{figure}[h]
	\begin{center}
		\includegraphics
		[width=0.35\textwidth]{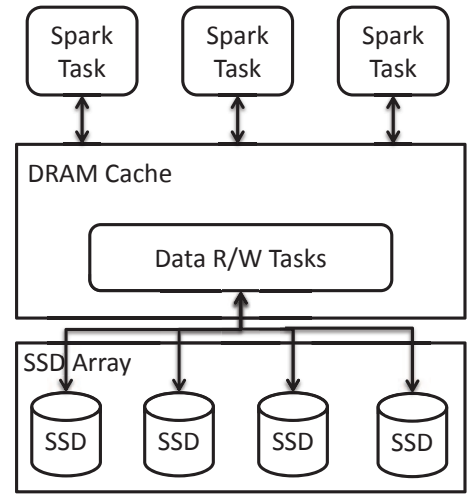}
		\vspace*{-0.05in}
		\caption{\small Multi-tier storage system consisting of DRAM and SSD.}
		\label{SSD_Spark}
		\vspace{-0.1in}
	\end{center}
\end{figure}

Spark takes DRAM as caches in its in-memory computation. Although DRAM has a much higher bandwidth and lower latency compared with HDD in data communication, its capacity is often limited due to the high cost of DRAM as well as its high power consumption~\cite{Badam:2011:SHS:1972457.1972479}. It can significantly constrain large-scale data applications from gaining high in-memory
hit-rates that is essential for high-performance on Spark. The new emerging storage devices in recent years give us a chance to alleviate it in the following ways:

\emph{SSD-based In-memory Computing}. Solid-State Disk (SSD) is a new storage device that provides much higher access speed than traditional HDD. Instead of using HDD, one approach is to adopt SSD as persistent storage by setting up a multi-tier storage system as illustrated in Figure~\ref{SSD_Spark}. In comparison to HDD, the data movement between memory and SSD is much faster. We can improve Spark performance by spilling RDDs to SSD when the memory cache is full. By using SSDs, there can be up to $10\times$ performance improvement over HDD-based caching approach for Spark~\cite{SSD_caching_DataBricks}.

\emph{NVM-based In-memory Computing}. Compared to DRAM, the latency of SSD is still very large (i.e., about $500\times$ slower than DRAM) although it is much faster than HDD~\cite{8310322}. Emerging Non-Volatile Memory (NVM), such as PCM, STT-RAM and ReRAM, is considered as an alternative to SSD~\cite{Liu:2017:HCC:3079079.3079089} due to its much lower latency and higher bandwidth than SSD. We can integrate DRAM, NVM and SSD to establish a multi-tier caching system by first caching the data in DRAM, or putting into NVM when DRAM is full, or in the SSD when both DRAM and SSD are full.

\section{Processor Supporting Layer}
\label{Processor_as_a_Support}

Since the limited performance and energy efficiency of general-purpose CPUs have impeded the performance scaling of conventional data centers, it becomes more and more popular to deploy accelerators in data centers, such as GPU and FPGA. Therefore, accelerator-based heterogeneous machine has become a promising basic block of modern data center to achieve further performance and efficiency.
In this section, we firstly provide a summary of Spark systems integrating with GPU to accelerate the computing task. Second, we make a survey of Spark systems with FPGA. 

\subsection{GPGPU}
GPU has been widely integrated into modern datacenter for its better performance and higher energy efficiency over CPU. However, the modern computing framework like Spark cannot directly leverage GPU to accelerate its computing task. Several related projects reach out to fill the gap.

\emph{1).HeteroSpark.} Li et al.~\cite{heterospark_nas15} present an novel GPU-enabled Spark \emph{HeteroSpark} which leverages the compute power of GPUs and CPUs to accelerate machine learning applications. The proposed GPU-enabled Spark provides a plug-n-play design so that the current Spark programmer can leverage GPU computing power without needing any knowledge about GPU.

\emph{2).Vispark.} Choi et al.~\cite{Vispark_isldav15} propose an extension of Spark called \emph{Vispark}, which leverages GPUs to accelerate array-based scientific computing and  processing applications. In particular, Vispark introduces VRDD (Vispark Resilient Distributed Dataset) for handling the array data on the GPU so that GPU computing abilities can be fully utilized.

\emph{3).Exploring GPU Acceleration of Apache Spark.} Manzi et al.~\cite{adobe_icce16} explore the possibilities and benefits of offloading the computing task of Spark to GPUs. In particular, the non-shuffling computing tasks can be computed on GPU and then the computation time is significantly reduced. The experimental result shows that the performance of K-Means clustering application was optimized by 17X. Its implementation is publicly available (https://github.com/adobe-research/spark-gpu).

\emph{4).Columnar RDD.} Ishizaki~\cite{spark_gpu_columnar_ibm} proposes one prototype which stores the inner data in a columnar RDD, compared with the conventional row-major RDD, since the columnar layout is much easier to benefit from using GPU and SIMD-enabled CPU. Therefore, the performance of the applicatin logistic regression is improved by 3.15X.

\subsection{FPGA}
FPGA is integrated into the computing framework Spark to accelerate inner computing task. In particular, there are two related projects: FPGA-enabled Spark and Blaze.

\emph{1).FPGA-enabled Spark~\cite{spark_fpga_hotcloud16}.} It explores how to efficiently integrate FPGAs into big-data computing framework Spark. In particular, it designs and deploys an FPGA-enabled Spark cluster, where one representative application next-generation DNA sequencing is accelerated with two key technologies. The first one is that they design one efficient mechanism to efficiently harness FPGA in JVM so that the JVM-FPGA communication (via PCIe) overhead is alleviated. The other one is that one FPGA-as-a-Service (FaaS) framework is proposed where FPGAs are shared among multiple CPU threads. Therefore, the computing abilities of FPGAs can be fully utilized and then the total execution time is significantly reduced.

\emph{2).Blaze~\cite{fpga_dacecenter_dac16}.} It provides a high-level programming interface (e.g., Java) to Spark and automatically leverages the accelerators (e.g., FPGA and GPU) in the heterogeneous cluster to speedup the computing task without the interference of programmer. In other words, each accelerator is abstracted as the subroutine for Spark task, which can be executed on local accelerator when it is available. Therefore, the computation time can be significantly reduced. Otherwise, the task will be executed on CPU.

\section{Data Management Layer}
\label{Data_Management_as_a_Support}

In the age of Big Data, data is generally stored and managed in distributed filesystems or databases. This sections gives a survey of widely used data storage and management systems for Spark.

\subsection{Distributed File Systems}

\emph{1). Hadoop Distributed File System (HDFS).}
Hadoop Distributed File System (HDFS) is proposed to be deployed on low-cost commodity hardware. It is highly scalable and fault-tolerant, enabling it to run on a cluster includes hundreds or thousands of nodes where the hardware failure is normal.
It takes a master-slave architecture, which contains a master called \emph{NameNode} to manage the file system namespace and regulating access to files by users, and a number of slaves called \emph{DataNodes} each located at a machine for storing the data. Data uploaded into HDFS are partitioned into plenty of blocks with fixed size (e.g., $64$ MB per data block) and the NameNode dispatched the data blocks to different DataNodes that store and manage the data assigned to them. To improve data reliability, it replicates each data block three times (the replicator is $3$ by default and users can change it) and stores each replica in a different rack.
HDFS data access has been originally supported by Spark with its provided native interface\footnote{Spark provides users the \emph{'spark-submit'} script to launch applications, which supports hdfs.}, which enables Spark applications to read/write data from/to HDFS directly.

\emph{2). Ceph.} The centralized nature inherent in the client/server model has testified a important barrier to scalable performance. Ceph~\cite{Weil:2006:CSH:1298455.1298485} is a distributed file system which offers high performance and dependability while promising unprecedented expansibility. Ceph uses generating functions replacing file allocation tables  to decouple the operations of data and metadata. Ceph is allowed to distribute the complexity around data access, update sequence, duplication and dependability, fault detection, and resume by using the intelligence in OSDs. Ceph uses a highly adaptive distributed metadata cluster architecture that greatly enhances the scalability of metadata access and the scalability of the whole system.

\emph{3). Alluxio.} With the rapid growth of today's big data, storage and networking pose the most challenging bottlenecks since data writes can become network or disk binding, especially when duplication is responsible for fault-tolerance. Alluxio~\cite{Alluxio}, used to be considered as Tachyon, is a fault-tolerant, memory-centric virtual distributed file system that can address the bottleneck. It enables reliable operation of memory speed and data sharing between different applications and cluster computing frameworks. To obtain high throughput writes without impairing fault-tolerance, Alluxio leverages the notion of lineage~\cite{Bose:2005:LRS:1057977.1057978} to recover the lost output by re-implementing output tasks, without the need of replicating the data.
With Alluxio, users can do transformations and explorations on large datasets in memory for high performance while enjoying its high data reliability.

\begin{figure}[h]
	\begin{center}
		\includegraphics
		[width=0.4\textwidth]{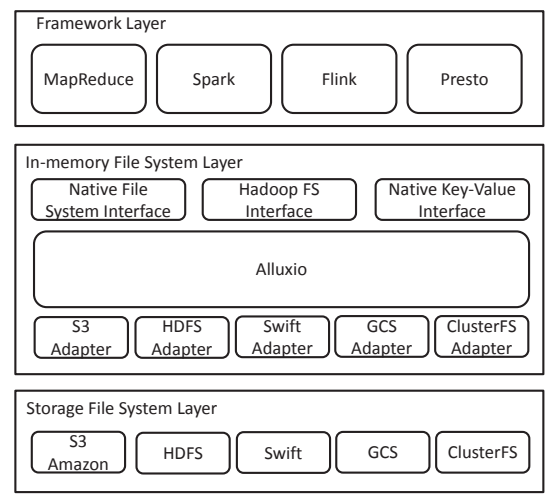}
		\vspace*{-0.05in}
		\caption{\small The Alluxio architecture.}
		\label{Aullxio_architecture}
		\vspace{-0.1in}
	\end{center}
\end{figure}

Figure~\ref{Aullxio_architecture} illustrates the memory-centric architecture of Alluxio. It manages data access and fast storage for user applications and computing frameworks by unifying the computing frameworks (e.g., MapReduce, Spark and Flink), and traditional storage systems (e.g., Amazon S3, Apache HDFS and OpenStack Swift), which facilitates data sharing and locality between jobs no matter whether they are running on the same computing system. It serves as a unifying platform for various data sources and computing systems. There are two key functional layers for Aullxio: lineage and persistence. The lineage layer offers high throughput I/O and tracks the information for tasks which produced a specific output. In contrast, the persistent layer materializes data into storage, which is mainly used for checkpoints. Aullxio employs a stand master-slave architecture. That master mainly manages the global metadata of the entire system, tracks lineage information and interacts with a cluster resource manager to distribute resources for recalculation. The slaves manage local storage resources allocated to Alluxio, and storing data and serving requests from users.


\subsection{Cloud Data Storage Services}
Cloud storage system is able to be typically viewed as a network of distributed data centers that provides storage service to users for storing data by using cloud computing techniques such as virtualization. It often stores the same data redundantly at different locations for high data availability, which is transparent to users. The cloud storage service can be accessed through a co-located cloud computer service, an application programming interfaces (API) or by applications that use the API~\cite{Cloud_storage}. There are two popular cloud storage services: Amazon S3 and Microsft Azure.

\emph{1). Amazon Simple Storage Service (S3)}. Amazon S3 is a web-based storage service that allows the user to store and fetch data at any time and any place through web services interfaces such as REST-style HTTP interface, SOSP interface and BitTorrent protocol~\cite{Amazon_S3}. It charges users for on-demand storage, requests and data transfers.

The data in Amazon S3 is managed as objects with an object storage architecture, which is opposed to file systems that manage data as a file hierarchy. Objects are organized into \emph{buckets}, each of which is owned by an AWS account. Users can identify objects within each bucket by a unique, user-assigned key.

Spark's file interface can allow users to access data in Amazon S3 by specifying a path in S3 as input through the same URI formats\footnote{The form of URI is: \emph{s3n://$<$bucket$>$/path}.} that are supported for Hadoop~\cite{AmazonS3}. However, the storage of Spark dataframe in Amazon S3 is not natively supported by Spark. Regarding this, users can utilize a spark s3 connector library~\cite{spark_S3} for uploading dataframes to Amazon S3.

\emph{2). Microsft Azure Blob Storage (WASB)}. Azure Blob storage (WASB)~\cite{Azure_Storage} is a cloud service for users to store and fetch any amount of unstructured data like text and binary data, in the form of Binary Large Objects (BLOBs). Three types of blobs are supported, namely, block blobs, append blobs and page blobs. Block blobs are suitable for storing and streaming cloud objects. Append blobs are optimized for append operations. In contrast, page blobs are improved to represent IaaS disks and support random writes. Multiple Blobs are grouped into a container and a user storage account can have any number of containers. The stored data can be accessed via HTTP, HTTPS, or REST API.

Spark is compatible with WASB, enabling the data stored in WASB to be directly accessed and processed by Spark via specifying an URI of the format \emph{`wasb://path'} that represents the path where the data is located.

\subsection{Distributed Database Systems}
\emph{1). Hbase.} Apache Hbase~\cite{HBase} is an open-source implementation of Google's BigTable~\cite{Chang:2006:BDS:1267308.1267323}, which is a distributed key-value database with the features of data compression, in-memory operation and bloom filters on a per-column basis. It runs on top of Hadoop that leverages the high scalability of HDFS and strong batch processing capabilities of MapReduce to enable massive data analysis, and provides real-time data access with the speed of a key/value store for individual record query.

It is a column-oriented key-value database that each table is stored as a multidimensional sparse map, having a timestamp for each cell tagged by column family and column name. A cell value can be identified and retrieved by specifying (Table Id, Row Key, Column-Family:Column, Timestamp). A Hbase table consists of regions, each of which is defined by a startKey and endKey. Except for parent column families being fixed in a schema, users can add columns to tables on-the-fly. All table accesses are achieved by the primary key through the Java API, REST, Avro or Thrift gateway APIs.

There are a number of libraries and tools emerged that enable Spark to interact with HBase. \emph{Spark-HBase Connector}~\cite{Spark_HBase_Connector} is such a library that provides a simple and elegant API for users' Spark applications to connect to HBase for reading and writing data. To enable native and optimized SQL access to HBase data via SparkSQL/Dataframe interfaces, a tool called \emph{Spark-SQL-on-HBase}~\cite{Spark-SQL-on-HBase} is developed by Huawei. Moreover, for efficient scanning, joining and mutating HBase tables to and from RDDs in a spark environment, there is a generic extension of spark module called \emph{spark-on-hbase}~\cite{spark-on-hbase} developed.

\emph{2). Dynamo.} Amazon Dynamo~\cite{DeCandia:2007:DAH:1323293.1294281} is a decentralized distributed key-value storage system with high scalability and availability for Amazon's applications. It has characteristics of both databases and distributed hash tables (DHTs)~\cite{Distributed_hash_table}. It is built to control the state of Amazon's application programs which require high reliability over the trade-offs between consistency, availability, cost-effectiveness and performance. Several Amazon e-commerce services only need primary-key access to a data store, such as shopping carts, customer preferences and sales rank. For these services, it caused inefficiencies and limited size and availability by using relational databases. In comparison, Dynamo is able to fulfill these requirements by providing a simple primary-key only interface.

Dynamo leverages a number of efficient optimization techniques to achieve high performance. It first uses a variant of consistent hashing to divide and replicate data across machines for overcoming the inhomogeneous data and workload distribution problem. Secondly, the technology is similar to arbitration and decentralized replication synchronization protocols to ensure data consistency during the update. Thirdly, it employs a gossip-style membership protocol that enables each node in the system to learn about the
arrival (or departure) of other nodes for the decentralized failure detection.

\emph{3). DynamoDB.} Amazon DynamoDB~\cite{Amazon_DynamoDB} is a new fast, high reliability, cost-effective NoSQL database service designed for Internet applications. It is based on strong distributed systems principles and data models of Dynamo. In contrast to Dynamo that requires users to run and manage the system by themselves, DynamoDB is a fully managed service that frees users from the headaches of complex installation and configuration operations. It is built on Solid State Drives (SSD) which offers fast and foreseeable performance with very low latency at any scale. It enables users to create a database table that can store and fetch any amount of data through the ability to disperse data and traffic to a sufficient number of machines to automatically process requests for any level of demand.

Medium company~\cite{Medium} creates a library called \emph{Spark-DynamoDB}~\cite{DynamoDB_Spark} that provides DynamoDB data access for Spark. It enables to read an DynamoDB table as a Spark DataFrame, and allows users to run SQL quries against DynamoDB tables directly with SparkSQL.

\emph{4). Cassandra.} Apache Cassandra~\cite{Cassandra} is a highly scalable, distributed structured key-value storage system designed to deal with large-scale data on top of hundreds or thousands of commodity servers. It is open sourced by Facebook in 2008 and has been widely deployed by many famous companies.

Cassandra integrates together the data model from Google's BigTable~\cite{Chang:2006:BDS:1267308.1267323} and distributed architectures of Amazon's Dynamo~\cite{DeCandia:2007:DAH:1323293.1294281}, making it eventually consistent like Dynamo and having a columnFamily-based data model like BigTable. Three basic database operations are supported with APIs: \textbf{insert}\emph{(table, key, rowMutation)}, \textbf{get}\emph{(table, key, columnName)} and \textbf{delete}\emph{(table, key, columnName)}. There are four main characteristics~\cite{Apache_Cassandra} for Cassandra. First, it is decentralized so that every node in the cluster plays the same role without introducing a single fault point of the master. Second, it is highly scalable that read/write throughput both increase linearly as the increasement of new machines and there is no downtime to applications. Third, each data is replicated automatically on multiple machines for fault tolerance and the failure is addressed without shutdown time. Finally, it offers a adjustable level of consistency, allowing the user to balance the tradeoff between read and write for different circumstances.

To enable the connection of Spark applicaitons to Cassandra, a \emph{Spark Cassandra Connector}~\cite{Spark_Cassandra_Connector} is developed and released openly by DataStax company.  It exposes Cassandra tables as Spark RDDs and can save RDDs back to Cassandra with an implicit \emph{saveToCassandra} call. Moreover, to provide the python support of pySpark~\cite{Python_Spark}, there is a module called \emph{pyspark-cassandra}~\cite{PySpark_Cassandra} built on top of \emph{Spark Cassandra Connector}.

\begin{table*}[htbp]\scriptsize
\begin{center}
\hspace*{-0.15in}
\begin{tabular}{|c|c|c|c|c|c|}
\hline
\textbf{Storage} & \textbf{System Type} & \textbf{Supported Layer} & \textbf{Data Model} & \textbf{Spark Query Interface} & \textbf{License} \\
\hline\hline
HDFS & Distributed File System & In Memory, In Disk & Document-Oriented Store & Low-Level API & Open source- Apache \\
\hline
Ceph & Distributed File System & In Disk & Document-Oriented Store & Low-Level API & Open source- LGPL \\
\hline
Alluxio & Distributed File System & In Memory, In Disk & Document-Oriented Store & Low-Level API & Open source- Apache \\
\hline
Amazon S3 & Cloud Storage System & In Disk & Object Store & Low-Level API & Commercial \\
\hline
Microsoft WASB & Cloud Storage System & In Disk & Object Store & Low-Level API & Commercial \\
\hline
Hbase & Distributed Database& In Disk & Key-Value Store & SparkSQL, Low-Level API& Open source- Apache \\
\hline
DynamoDB & Distributed Database& In Disk & Key-Value Store & SparkSQL, Low-Level API& Commercial \\
\hline
Cassandra & Distributed Database& In Memory, In Disk & Key-Value Store & SparkSQL, Low-Level API& Open source- Apache \\
\hline
\end{tabular}
\caption{\small The comparison of different storage systems.}
\label{Storage_systems}
\end{center}
\vspace*{-0.1in}
\end{table*}

\subsection{Comparison}


Table~\ref{Storage_systems} shows the comparison of different storage systems supported by Spark. We summarize them in different ways, including the type of storage systems they belong to, the storage places where it supports to store the data, the data storing model, the data accessing interface and the licence. Similar to Hadoop, Spark has a wide range support for various typed storage systems via its provided low-level APIs or SparkSQL, which is crucial to keep the generality of Spark from the data storage perspective.  Like Spark's in-memory computation, the in-memory data caching/storing is also very important for achieving high performance. HDFS, Alluxio and Cassandra can support in-memory and in-disk data storage manners, making them become most popular and widely used for many big data applications.

\section{Data Processing Layer}
\label{Processing_as_a_Support}

As a general-purpose framework, Spark supports a variety of data computation, including Streaming Processing, Graph Processing, OLTP and OLAP Queries Processing, and Approximate Processing. This section discusses about research efforts on them.

\subsection{Streaming Processing}
Spark Streaming allows data engineers and data scientists to process real-time data from various sources like Kafka, Flume, and Amazon Kinesis.
Spark is built upon the model of data parallel computation. It provides reliable processing of live streaming data. Spark streaming transforms streaming computation into a series of deterministic micro-batch computations, which are then executed using Spark's distributed processing framework.
The key abstraction is a Discretized Stream~\cite{Discretized_Streams} which represents a stream of data divided into small batches.
The way Spark Streaming works is that it divides the live stream of data into batches (called microbatches) of a pre-defined interval (N seconds) and then treats each batch of data as Resilient Distributed Datasets (RDDs)~\cite{RDD}. This allows Spark Streaming to seamlessly integrate with any other Spark components like MLlib and Spark SQL.
Due to the popularity of spark streaming, research efforts are devoted on further improving it. Das et al.~\cite{das2014adaptive} study the relationships among batch size, system throughput and end-to-end latency.

There are also efforts to extend spark streaming framework.

\emph{1). Complex Event Processing.}
Complex event processing (CEP) is a type of event stream processing that combines data from multiple sources to identify patterns and complex relationships across various events.
CEP system helps identify opportunities and threats across many data sources and provides real-time alerts to act on them.
Over the last decades, CEP systems have been successfully applied in a variety of domains such as recommendation, stock market monitoring, and health-care.
There are two open-source projects on building CEP system on Spark.
Decision CEP engine~\cite{Decision} is a Complex Event Processing platform built on Spark Streaming.
It is the result of combining the power of Spark Streaming as a continuous computing framework and Siddhi CEP engine as complex event processing engine.
Spark-cep~\cite{Sparkcep} is another stream processing engine built on top of Spark supporting continuous query language. Comparing to the existing Spark Streaming query engines, it supports more efficient windowed aggregation and ``Insert Into" query.

\emph{2). Streaming Data Mining.}
In this big data era, the growing of streaming data motivates the fields of streaming data mining. There are typically two reasons behind the need of evolving from traditional data mining approach.
First, streaming data has, in principle, no volume limit, and hence it is often impossible to fit the entire training dataset into main memory.
Second, the statistics or characteristics of incoming data are continuously evolving, which requires a continuously re-training and evolving. Those challenges make the traditional offline model approach no longer fit.
To this end, open-sourced distributed streaming data mining platforms, such as SOMOA~\cite{morales2015samoa} and StreamDM~\cite{streamDM} are proposed and have attracted many attentions.
Typically, StreamDM~\cite{streamDM,7395869} uses Spark Streaming as the provider of streaming data.
A list of data mining libraries are supported such as SGD Learner and Perception.

\subsection{Graph Processing}
Many practical computing problems concern large graphs.
As graph problems grow larger in scale and more ambitious in their complexity, they easily outgrow the computation and memory capacities. To this end, distributed graph processing frameworks such as GraphX~\cite{GraphX} are proposed.
GraphX is a library on top of Spark by encoding graphs as collections and then expressing
the GraphX API on top of standard dataflow operators.
In GraphX, a number of optimization strategies are developed, and we briefly mention a few here.
\begin{itemize}
	\item
	GraphX includes a range of built-in partitioning functions. The vertex collection is hash-partitioned by vertex ids. The edge collection is horizontally partitioned by a user-defined function, supporting vertex-cut partitioning. A routing table is co-partitioned with the vertex collection.
	\item
	For maximal index reuse, subgraph operations produce subgraphs that share the full graph indexes, and use bitmasks to indicate which elements are included.
	\item
	In order to reduce join operation, GraphX uses JVM bytecode analysis to determine what properties a user-defined function accesses. With a not-yet materialized triplets view, and only one property accessed GraphX will use a two-way join. With no properties accessed, GraphX can eliminate the join completely. 	
\end{itemize}

In contrast to many specialized graph processing system such as Pregel~\cite{Malewicz:2010:PSL:1807167.1807184}, PowerGraph~\cite{Gonzalez:2012:PDG:2387880.2387883}, GraphX is closely integrated into modern general-purpose distributed dataflow system (i.e., Spark).
This approach avoids the need of composing multiple systems which increases complexity for a integrated analytics pipelines, and reduces unnecessary data movement and duplication.
Furthermore, it naturally inherited the efficient fault tolerant feature from Spark, which is usually overlooked in specialized graph processing framework. The evaluation also shows that GraphX is comparable to or faster than specialized graph processing systems.

\subsection{OLTP and OLAP Queries Processing}

Hybrid Transaction/Analytical Processing (HTAP) systems support both OLTP and OLAP queries by storing data in dual formats but need to be used alongside a streaming engine to support streaming processing. SnappyData~\cite{Ramnarayan:2016:SHT:2882903.2899408} enable streaming, transactions and interactive analytics in a single unifying system and exploit AQP techniques and a variety of data synopses at true interactive speeds. SnappyData consists of a deep integration of Apache Spark and GemFire. An operational of in-memory data storage is combined with Spark's computational model. When Spark executes tasks in a partitioned manner, it keeps all available CPU cores busy. Spark's API are extended to unified API for OLAP, OLTP, and streaming.

\begin{figure}[h]
	\begin{center}
		\includegraphics
		[width=0.45\textwidth]{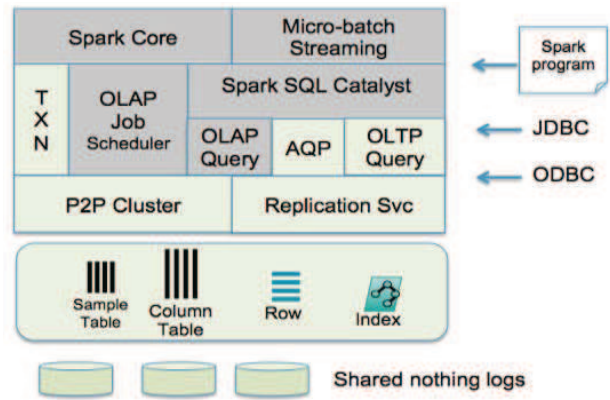}
		\caption{\small The Core Components of SnappyData~\cite{Ramnarayan:2016:SHT:2882903.2899408}.}
		\label{SnappyData_core_components}
	\end{center}
\end{figure}

As shown in Figure~\ref{SnappyData_core_components}, Spark's original components are highlighted in gray. The storage layer is primarily in-memory and manages data in either row or column formats. The OLAP scheduler and job server coordinate all OLAP and Spark jobs and all OLTP operations are routed to appropriate partitions without any scheduling. A P2P cluster membership service is utilized to ensure view consistency and virtual synchrony.

\subsection{Approximate Processing}


Modern data analytics applications demand near real-time response rates. However, getting exact answer from extreme large size of data takes long response time, which is sometimes unacceptable to the end users. Besides using additional resources (i.e., memory and CPU) to decrease data processing time, approximate processing provides faster query response by reducing the amount of work need to perform through techniques such as sampling or online aggregation. It has been widely observed that users can accept some inaccurate answers which come quickly, especially for exploratory queries.

\emph{1). Approximate Query Processing}. In practice, having a low response time is crucial for many applications such as web-based interactive query workloads. To achieve that, Sameer et al.~\cite{BlinkDB} proposed a approximate query processing system called BlinkDB atop of Shark and Spark, based on the distributed sampling. It can return the query result for a large queries of $17$ full data terabytes within $2$ seconds while keeping meaningful error bounds relative to the answer with $ 90-98\%$.  The strength of BlinkDB comes from two key ideas: (1) an adaptive optimization framework that builds and maintains a set of multi-dimensional samples from original data over time, and (2) a dynamic sample selection strategy that selects an appropriately sized sample based on a query's accuracy and/or response time requirements. Moreover, in order to evaluate the accuracy of BlinkDB, Agarwal et al.~\cite{Agarwal2014Knowing} proposed an effective error estimation approach by extending the prior diagnostic algorithm~\cite{Kleiner:2013:GBP:2487575.2487650} to detect when bootstrap based error estimates are unreliable.

Considering that the join operation is a key building block for any database system, Quoc et al.~\cite{2018arXiv180505874L} proposed a new join operator called APPOXJOIN that approximates distributed join computations on top of Spark by interweaving Bloom filter sketching and stratified sampling. It first uses a Bloom filter to avoid shuffling non-joinable data and next leverages the stratified sampling approach to get a representative sample of the join output.

\emph{2).Approximate Streaming Processing.} Unlike the batch analytics where the input data keep unchanged during the sampling process, the data for streaming analytics is changing over time. Quoc et al.~\cite{2017arXiv170902946L} shows that the traditional batch-oriented approximate computing are not well-suited for streaming analytics. To address it, they proposed a streaming analytics system called STREAMAPROX by designing an online stratified reservoir sampling algorithm to generate approximate output with rigorous error bounds. It implements STREAMAPROX on Apache Spark Streaming and experimental results show that there can be a speedup of $1.1\times-2.4\times$ while keeping the same accuracy level over the baseline of Spark-based approximate computing system leveraging the existing sampling modules in Apache Spark.

\emph{3).Approximate Incremental Processing.} Incremental processing refers to a data computation that is incrementally scheduled by repeatedly involving the same application logic or algorithm logic over an input data that differs slightly from previous invocation~\cite{Gunda:2010:NAM:1924943.1924949} so as to avoid recomputing everything from scratch. Like approximate computation, it works over a subset of data items but differ in their choosing means. Krishnan et al.~\cite{Krishnan:2016:IDA:2872427.2883026} observe that the two paradigms are complementary and proposed a new paradigm called
approximate incremental processing that leverages the approximation and incremental techniques in order for a low-latency execution. They designed an online stratified sampling algorithm by leveraging self-adjusting computation to generate an incrementally updated approximate output with bounded error and implemented it in Apache Spark Streaming by proposing a system called INCAPPROX. The experimental evaluation shows that benefits of INCAPPROX equipping with incremental and approximate computing.

\section{High-level Language Layer}
\label{Language_as_a_Support}

Spark is written in Scala~\cite{Scala_Language}, which is an object-oriented, functional programming language running on a Java virtual machine that can call Java libraries directly in Scala code and vice versa. Thus, it natively supports the Spark programming with Scala and Java by default. However, some users might be unfamiliar with Scala and Java but are skilled in other alternative languages like Python and R. Moreover, Spark programming is still a complex and heavy work especially for users that are not familiar with Spark framework. Thereby, having a high-level language like SQL declarative language on top of Spark is crucial for users to express their tasks while leave all of the complicated execution optimization details to the backend Spark engine, which alleviates users' programming burdens significantly. In the following section, we discuss about research efforts that have been proposed to address these problems.

\subsection{R and Python High-level Languages Support}
\emph{1). SparkR.} In the numeric analysis and machine learning domains, R~\cite{R_programmingLanguage} is a popular programming language widely used by data scientists for statistical computing and data analysis. SparkR~\cite{venkataramansparkr,SparkR} is a light-weight frontend system that incorporates R into Spark and enables R programmers to perform large-scale data analysis from the R shell. It extends the single machine implementation of R to the distributed data frame implementation on top of Spark for large datasets.
The implementation of SparkR is on the basis of Spark's parallel DataFrame abstraction~\cite{Spark_SQL}. It supports all Spark DataFrame analytical operations and functions including aggregation, filtering, grouping, summary statistics, and mixing-in SQL queries.



\emph{2). PySpark.} PySpark~\cite{PySpark} is the Python API for Spark, which exposes the Spark programming model to Python. It allows users to write Spark applications in Python. There are a few differences between PySpark and Spark Scala APIs. First, Python is a dynamically typed language so that the RDDs of PySpark have the capability to store objects of multiple types. Second, the RDDs of PySpark support the same functions as that of Scala APIs but leverage Python functions and return Python collection types. Third, PySpark supports anonymous functions that can be passed as arguments to the PySpark API by using Python's lambda functions.


\subsection{SQL-like Programming Language and System}

\emph{1). Shark.} Apache Shark~\cite{Engle:2012:SFD:2213836.2213934,Shark} is the first SQL-on-Spark effort. It is built on top of Hive codebase and uses Spark as the backend engine. It leverages the Hive query compiler (HiveQL Parser) to parse a HiveQL query and generate an abstract syntax tree followed by turning it into the logical plan and basic logical optimization. Shark then generates a physical plan of RDD operations and finally executes them in Spark system. A number of performance optimizations are considered. To reduce the large memory overhead of JVM, it implements a columnar memory store based on Spark's native memory store. A cost-based query optimizer is also implemented in Shark for choosing more efficient join order according to table and column statistics. To reduce the impact of garbage collection, Shark stores all columns of primitive types as JVM primitive arrays. Finally, Shark is completely compatible with Hive and HiveQL, but much faster than Hive, due to its inter-query caching of data in memory that eliminates the need to read/write repeatedly on disk. It can support more complex queries through User Defined Functions (UDFs) that are referenced by a HiveQL query.


\emph{2). Spark SQL.} As an evolution of SQL-on-Spark, Spark SQL~\cite{Spark_SQL} is the state-of-art new module of Spark that has replaced Shark in providing SQL-like interfaces. It is proposed and developed from ground-up to overcome the difficulty of performance optimization and maintenance of Shark resulting from inheriting a large, complicated Hive codebase. Compared to Shark, it adds two main capabilities. First, Spark SQL provides much tighter hybrid of relational and procedural processing. Second, it becomes easy to add composable rules, control code generation, and define extension points. It is compatible with Shark/Hive that supports all existing Hive data formats, user-defined functions (UDF) and the Hive metastore, while providing the state-of-the-art SQL performance.

\begin{figure}[h]
	\begin{center}
		\includegraphics
		[width=0.4\textwidth]{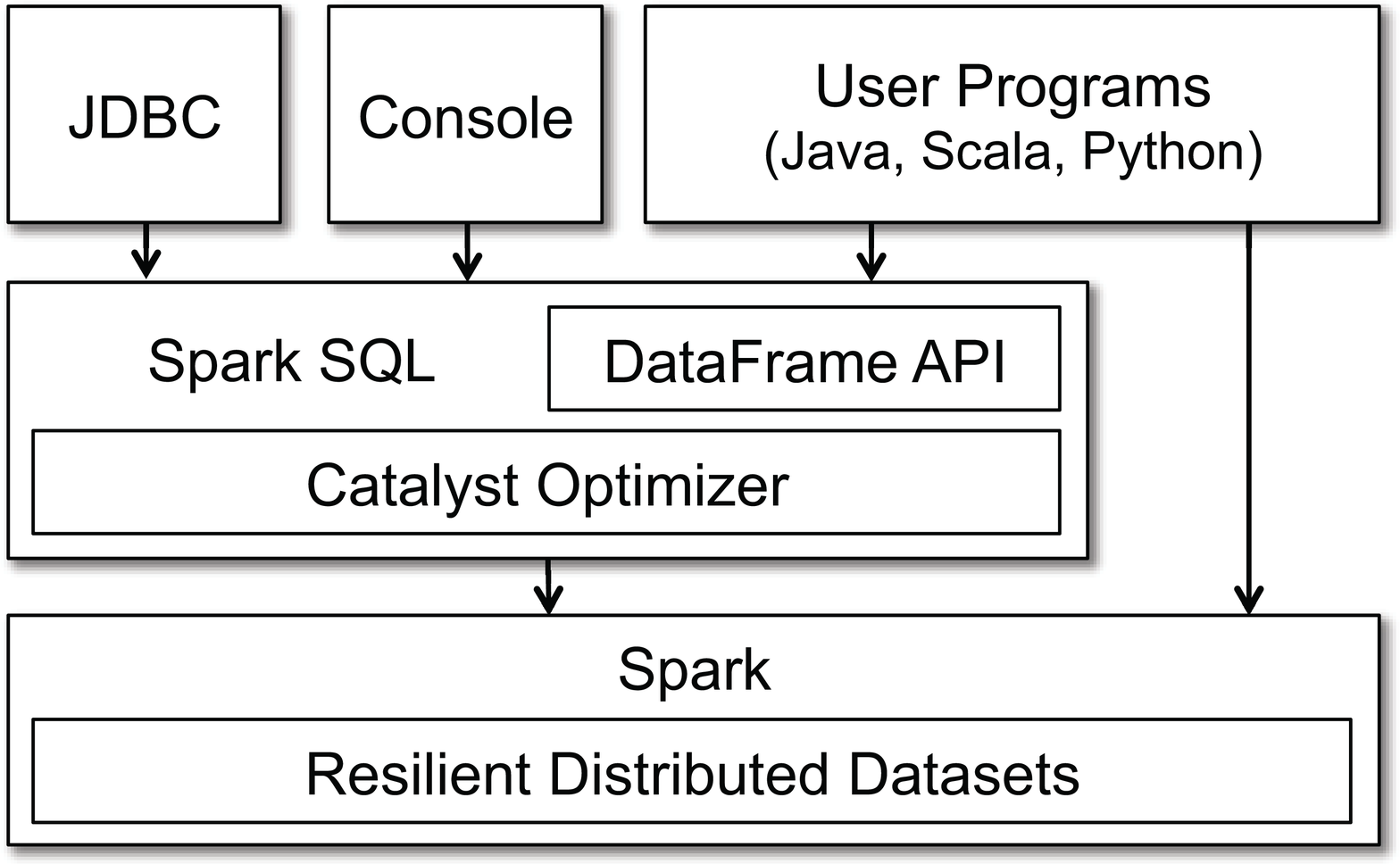}
		\vspace*{-0.05in}
		\caption{\small Interfaces to Spark SQL, and interaction with Spark.~\cite{Spark_SQL}}
		\label{SparkSQL_interface}
	\end{center}
\end{figure}

Figure~\ref{SparkSQL_interface} presents the programming interface to Spark SQL containing two main cores of DataFrame API and Catalyst Optimizer, and its interaction with Spark.  It exposes SQL interfaces through JDBC/ODBC, a command-line console, and the DataFrame API implemented in Spark's supported programming languages. The DataFrame is the main abstraction in Spark SQL's API. It is a distributed collections of records that can be operated with Spark's procedural API, or new relational APIs. The Catalyst, in contrast, is an extensible query optimizer based on functional programming constructs. It simplifies the addition of new optimization techniques and features to Spark SQL and enables users to extend the optimizer for their application needs.

\emph{3). Hive/HiveQL.} Apache Hive~\cite{Hive} is an open-source data warehousing solution built on top of Hadoop by the Facebook Data Infrastructure Team. It aims to incorporate the classical relational database notion as well as high-level SQL language to the unstructured environment of Hadoop for those users who were not familiar with map-reduce. There is a mechanism inside Hive that can project the structure of table onto the data stored in HDFS and enable data queries using a SQL-like declarative language called HiveQL, which contains its own type system with support for tables, collections and nested compositions of the same and data definition language (DDL).  Hive compiles the SQL-like query expressed in HiveQL into a directed acyclic graph (DAG) of map-reduce jobs that are executed in Hadoop. There is a metastore component inside Hive that stores metadata about the underlying table, which is specified during table creation and reused whenever the table is referenced in HiveQL. The DDL statements supported by HiveQL enable to create, drop and alter tables in a Hive database. Moreover, the data manipulation statements of HiveQL can be used to load data from external sources such as HBase and RCFile, and insert query results into Hive tables.

Hive has been widely used by many organizations/users for their applications~\cite{Hive_on_Spark_analysis}. However, the default backend execution engine for Hive is MapReduce, which is less powerful than Spark. Adding Spark as an alternative backend execution engine to Hive is thus an important way for Hive users to migrate the execution to Spark. It has been realized in the latest version of Hive~\cite{Apache_Hive}. Users can now run Hive on top of Spark by configuring its backend engine to Spark.

\begin{figure}[h]
	\begin{center}
		\includegraphics
		[width=0.45\textwidth]{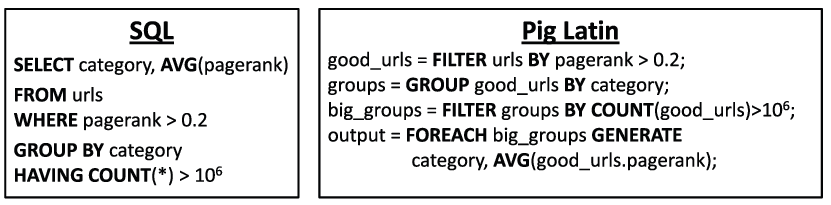}
		\caption{\small An example of SQL Query and its equivalent Pig Latin program.~\cite{Apache_Pig}}
		\label{PigLatin_Example}
	\end{center}
\end{figure}

\emph{4). Pig/Pig Latin.} Apache Pig~\cite{Apache_Pig} is an open source dataflow processing system developed by Yahoo!, which serves for experienced procedural programmers with the preference of map-reduce style programming over the pure declarative SQL-style programming in pursuit of more control over the execution plan. It consists of a execution engine and high-level data flow language called \emph{Pig Latin}~\cite{Olston:2008:PLN:1376616.1376726}, which is not declarative but enables the expression of a user's task using high-level declarative querying in the spirit of SQL and low-level procedural programming with MapReduce. Figure~\ref{PigLatin_Example} gives an example of SQL query and its equivalent Pig Latin program, which is a sequence of transformation steps each of which is carried out using SQL-like high-level primitives (e.g., filtering, grouping, and aggregation). Given a Pig Latin program, the Pig execution engine generates a logic query plan, compiles it into a DAG of MapReduce jobs, and finally submitted to Hadoop cluster for execution.

There are several important characteristics for Pig Latin in casual ad-hoc data analysis, including the support of a nested data model as well as a set of predefined and customizable user-defined functions (UDFs), and the ability of operating over plain files without any schema information. In Pig Latin, the basic data type is Atom (e.g., integer, double, and string). Multiple Automs can be combined into a Tuple and several Tuples can form a Bag. Map is a more complex data type supported by Pig Latin, which contains a key and a collection of data items that can be looked up with its associated key.

Like Hive, the default backend execution engine for Pig is MapReduce. To enable the execution of Pig jobs on Spark for performance improvement, there is a Pig-on-Spark project called Spork~\cite{Spork} that plugs in Spark as an execution engine for Pig. With Spork, users can choose Spark as the backend execution engine of the Pig framework optionally for their own applications.

\subsection{Comparison}

\begin{table*}[htbp]\scriptsize
\begin{center}
\hspace*{-0.15in}
\begin{tabular}{|c||c|c|c|c|c|}
\hline
\textbf{System} & \textbf{Language Type} & \textbf{Data Model} & \textbf{UDF} & \textbf{Access Interface} & \textbf{MetaStore} \\
\hline\hline
SparkR & Dataflow, SQL-like & Nested & Supported & Command line, web, JDBC/ODBC server & Supported \\
\hline
PySpark & Dataflow,  SQL-like & Nested & Supported & Command line, web, JDBC/ODBC server & Supported \\
\hline
Shark & SQL-like & Nested & Supported & Command line & Supported \\
\hline
SparkSQL & SQL-like & Nested & Supported & Command line, web, JDBC/ODBC server & Supported \\
\hline
Hive &  SQL-like & Nested & Supported & Command line, web, JDBC/ODBC server & Supported \\
\hline
Pig & Dataflow & Nested & Supported & Command line & Not supported \\
\hline
\end{tabular}
\caption{\small The comparison of different programming language systems.}
\label{Language_System}
\end{center}
\vspace*{-0.1in}
\end{table*}


Table~\ref{Language_System} illustrates the comparison of different programming language systems used in Spark. To be compatible, it supports Hive and Pig by allowing users to replace the backend execution engine of MapReduce with Spark. To make the query efficient, Shark is first developed and later evolves to SparkSQL. Moroever, SparkR and PySpark are provided in Spark in order to support R and Python languages which are widely used by scientific users. Among these languages, the major differences lie in their supported language types. SparkR and PySpark can support Dataflow and SQL-like programming. In contrast, Shark, SparkSQL and Hive are SQL-like only languages, while Pig is a dataflow language.

\section{Application/Algorithm Layer}
\label{Application_as_a_Support}
As a general-purpose system, Spark has been widely used for various applications and algorithms. In this section, we first review the support of machine learning algorithms on Spark. Next we show the supported applications on Spark.

\subsection{Machine Learning Support on Spark}
\label{Machine_Learning_Support}

Machine learning is a powerful technique used to develop personalizations, recommendations and predictive insights in order for more diverse and more user-focused data products and services. Many machine learning algorithms involve lots of iterative computation in execution. Spark is an efficient in-memory computing system for iterative processing. In recent years, it attracts many interests from both academia and industry to build machine learning packages or systems on top of Spark. In this section, we discuss about research efforts on it.

\subsubsection{Machine Learning Library}

\emph{1). MLlib.} The largest and most active distributed machine learning library for Spark is MLlib~\cite{meng2015mllib,MLlib}. It consists of fast and scalable implementations of common machine learning algorithms and a variety of basic analytical utilities, low-level optimization primitives and higher-level pipeline APIs. It is a general machine learning library that provides algorithms for most use cases and meanwhile allows users to build upon and extend it for specialized use cases.

There are several core features for MLlib as follows. First, it implements a number of classic machine learning algorithms, including various linear models (e.g., SVMs, logistic regression, linear regression), naive Bayes, and ensembles of decision trees for classification and regression problems; alternating least squares for collaborative filtering; and k-means clustering and principal component analysis for clustering and dimensionality reduction; FP-growth for frequent pattern mining. Second, MLlib provides many optimizations for supporting efficient distributed learning and prediction. Third, It supports practical machine learning pipelines natively by using a package called spark.ml inside MLlib, which simplifies the development and tuning of multi-stage learning pipelines by providing a uniform set of high-level APIs. Lastly, there is a tight and seamless integration of MLlib with Spark's other components including Spark SQL, GraphX, Spark streaming and Spark core, bringing in high performance improvement and various functionality support for MLlib.

MLlib has many advantages, including simplicity, scalability, streamlined end-to-end and compatibility with Spark's other modules. It has been widely used in many real applications like marketing, advertising and fraud detection.

\emph{2). KeystoneML.} KeystoneML~\cite{7930005} is a framework for ML pipelines, written in Scala, from the UC Berkeley AMPLab designed to simplify the construction of large scale, end-to-end, machine learning pipelines with Apache Spark. It captures and optimizes the end-to-end large-scale machine learning applications for high-throughput training in a distributed environment with a high-level API~\cite{KeystoneML}. KeystoneML has several core features.  First, it allows users to specify end-to-end ML applications in a single system using high level logical operators. Second, it scales out dynamically as data volumes and problem complexity change. Finally, it Automatically optimizes these applications given a library of ML operators and the user's compute resources. KeystoneML is open source software and is being used in scientific applications in solar physics~\cite{2016AGUFMSH34A02J} and genomics~\cite{Synapse}.

\emph{3). Thunder.} Thunder~\cite{Thunder} is an open-source library developed by Freeman Lab~\cite{Freeman_Lab} for large-scale neural data analysis with Spark. It is written in Spark Python API (PySpark) for the use of robust numerical and scientific computing libraries (e.g., NumPy and SciPy), and offers the simplest front end for new users. Thunder provides a set of data structures and utilities for loading and saving data using a variety of input formats, classes for dealing with distributed spatial and temporal data, and modular functions for time series analysis,  processing, factorization, and model fitting~\cite{cunningham2014analyzing}. It can be used in a variety of domains including medical imaging, neuroscience, video processing, and geospatial and climate analysis.

\emph{4). ADAM.} ADAM~\cite{ADAM} is a library and parallel framework that enables to work with both aligned and unaligned genomic data using Apache Spark across cluster/cloud computing environments. ADAM provides competitive performance to optimized multi-threaded tools on a single node, while enabling scale out to clusters with more than a thousand cores. ADAM is built as a modular stack, which is different from traditional genomics tools. This stack architecture supports a wide range of data formats and optimizes query patterns without changing data structures. There are seven layers of the stack model from bottom to top: Physical Storage, Data Distribution, Materialized Data, Data Schema, Evidence Access, Presentation, Application~\cite{MassieADAM}. A ``narrow waisted'' layering model is developed for building similar scientific analysis systems to enforce data independence. This stack model separates computational patterns from the data model, and the data model from the serialized representation of the data on disk. They exploit smaller and less expensive computers, leading to a $63\%$ cost improvement and a $28\times$ improvement in read preprocessing pipeline latency~\cite{Nothaft2015Rethinking}.

\subsubsection{Machine Learning System}

\emph{1). MLBase.} The complexity of existing machine learning algorithms is so overwhelming that users often do not understand the trade-offs and challenges of parameterizing and picking up between different learning algorithms for achieving good performance. Moreover, existing distributed systems that support machine learning often require ML researchers to have a strong background in distributed systems and low-level primitives. All of these limits the wide use of machine learning technique for large scale data sets seriously. MLBase~\cite{kraska2013mlbase,talwalkar2012mlbase} is then proposed to address it as a platform.

\begin{figure}[h]
	\begin{center}
		\includegraphics
		[width=0.4\textwidth]{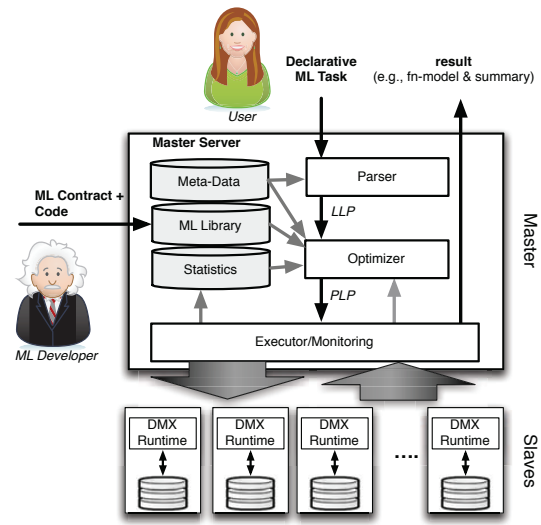}
		\vspace*{-0.05in}
		\caption{\small MLbase Architecture.~\cite{kraska2013mlbase}}
		\label{MLbase_architecture}
	\end{center}
\end{figure}

The architecture of MLBase is illustrated in Figure~\ref{MLbase_architecture}, which contains a single master and a set of slave nodes.  It provides a simple declarative way for users to express their requests with the provided declarative language and submit to the system. The master parses the request into a logical learning plan (LLP) describing the most general workflow to perform the request. The whole search space for the LLP can be too huge to be explored, since it generally involves the choices and combinations of different ML algorithms, algorithm parameters, featurization techniques, and data sub-sampling strategies, etc. There is an optimizer available to prune the search space of the LLP to get an optimized logical plan in a reasonable time. After that, MLBase converts the logical plan into a physical learning plan (PLP) making up of executable operations like filtering, mapping and joining.
Finally, the master dispatches these operations to the slave nodes for execution via MLBase runtime.

\emph{2). Sparkling Water.} H2O~\cite{H2O} is a fast, scalable, open-source, commercial machine learning system produced by \emph{H2O.ai Inc.}~\cite{H2O.ai} with the implementation of many common machine learning algorithms including generalized linear modeling (e.g., linear regression, logistic regression), Naive Bayes, principal components analysis and k-means clustering, as well as advanced machine learning algorithms like deep learning, distributed random forest and gradient boosting. It provides familiar programming interfaces like R, Python and Scala, and a graphical-user interface for the ease of use. To utilize the capabilities of Spark, Sparkling Water~\cite{sparkling-water} integrates H2O's machine learning engine with Spark transparently. It enables launching H2O on top of Spark and using H2O algorithms and H2O Flow UI inside the Spark cluster, providing an ideal machine learning platform for application developers.

Sparking Water is designed as a regular Spark application and launched inside a Spark executor spawned after submitting the application. It offers a method to initialize H2O services on each node of the Spark cluster. It enables data sharing between Spark and H2O with the support of transformation between different types of Spark RDDs and H2O's H2OFrame, and vice versa.

\emph{3). Splash.} Stochastic algorithms are efficient approaches to solving machine learning and optimization problems. Splash~\cite{Splash} is a framework for parallelizing stochastic algorithms on multi-node distributed systems, it consists of a programming interface and an execution engine. Users use programming interface to develop sequential stochastic algorithms and then the algorithm is automatically parallelized by a communication-efficient execution engine. Splash can be called in a distributed manner for constructing parallel algorithms by execution engine. In order to parallelize the algorithm, Splash converts a distributed processing task into a sequential processing task using distributed versions of averaging and reweighting. Reweighting scheme ensures the total weight processed by each thread is equal to the number of samples in the full sequence. This helps individual threads to generate nearly unbiased estimates of the full update. Using this approach, Splash automatically detects the best degree of parallelism for the algorithm. The experiments verify that Splash can yield orders-of-magnitude speedups over single-thread stochastic algorithms and over state-of-the-art batch algorithms.

\emph{4). Velox.} BDAS(Berkeley Data Analytics Stack) contained a data storage manager, a dataflow execution engine, a stream processor, a sampling engine, and various advanced analytics packages. However, BDAS lacked any means of actually serving data to end-users, and, there are many industrial users of the stack rolled their own solutions to model serving and management. Velox fills this gap. Velox~\cite{Crankshaw2014The} is a system for performing model serving and model maintenance at scale. It provides end-user applications and services with a low latency, intuitive interface to models, transforming the raw statistical models currently trained using existing offline large-scale compute frameworks into full-blown, end-to-end data products capable of recommending products, targeting advertisements, and personalizing web content.Velox consists of two primary architectural components: Velox model manager and Velox model predictor. Velox model manager orchestrates the computation and maintenance of a set of pre-declared machine learning models, incorporating feedback and new data, evaluating model performance, and retraining models as necessary.


\subsubsection{Deep Learning}
As a class of machine learning algorithms, \emph{Deep learning} has become very popular and been widely used in many fields like computer version, speech recognition, natural language processing and bioinformatics due to its many benefits: accuracy, efficiency and flexibility. There are a number of deep learning frameworks implemented on top of Spark, such as CaffeOnSpark~\cite{CaffeOnSpark}, DeepLearning4j~\cite{deeplearning4j}, and SparkNet~\cite{moritz2015sparknet}.

\begin{figure}[htp]
	\begin{center}\hspace*{-0.15in}
		\subfloat[\small ML Pipeline with multiple programs on separated clusters. ]{\label{fig:priceprobc1}\includegraphics[width=0.32\textwidth]{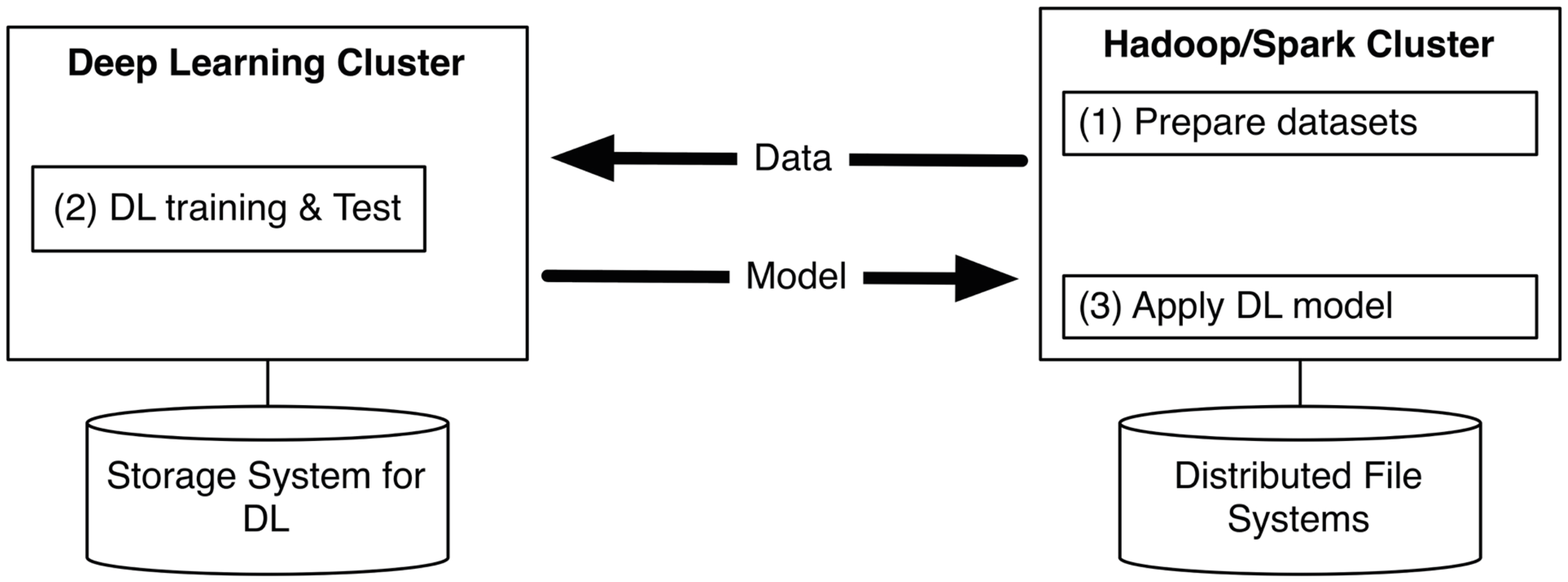}}\hspace*{0.05in}
		\begin{flushleft}
			
		\end{flushleft}
		\subfloat[\small ML Pipeline with single program on one cluster. ]{\label{fig:priceprobc1}\includegraphics[width=0.2\textwidth]{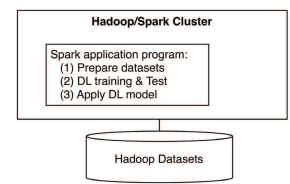}}
		\caption{\small Distributed deep learning computing model.~\cite{CaffeOnSpark_Introduction}}
		\label{Spark_based_ML}
	\end{center}
\end{figure}

\emph{1). CaffeOnSpark.} In many existing distributed deep learning, the model training and model usage are often separated, as the computing model shown in Figure~\ref{Spark_based_ML}(a). There is a big data processing cluster (e.g., Hadoop/Spark cluster) for application computation and a separated deep learning cluster for model training. To integrate the model training and model usage as a united system, it requires a large amount of data and model transferred between two separated clusters by creating multiple programs for a typical machine learning pipeline, which increases the system complexity and latency for end-to-end learning. In contrast, an alternative computing model, as illustrated in Figure~\ref{Spark_based_ML}(b), is to conduct the deep learning and data processing in the same cluster.

Caffe~\cite{jia2014caffe} is one of the most popular deep learning frameworks, which is developed in C++ with CUDA by Berkeley Vision and Learning Center (BVLC). According to the model of Figure~\ref{Spark_based_ML}(b), Yahoo extends Caffe to Spark framework by developing CaffeOnSpark~\cite{CaffeOnSpark_Introduction,CaffeOnSpark}, which enables distributed deep learning on a cluster of GPU and CPU machines. CaffeOnSpark is a Spark package for deep learning, as a complementary to non-deep learning libraries MLlib and Spark SQL.

\begin{figure}[h]
	\begin{center}
		\includegraphics
		[width=0.4\textwidth]{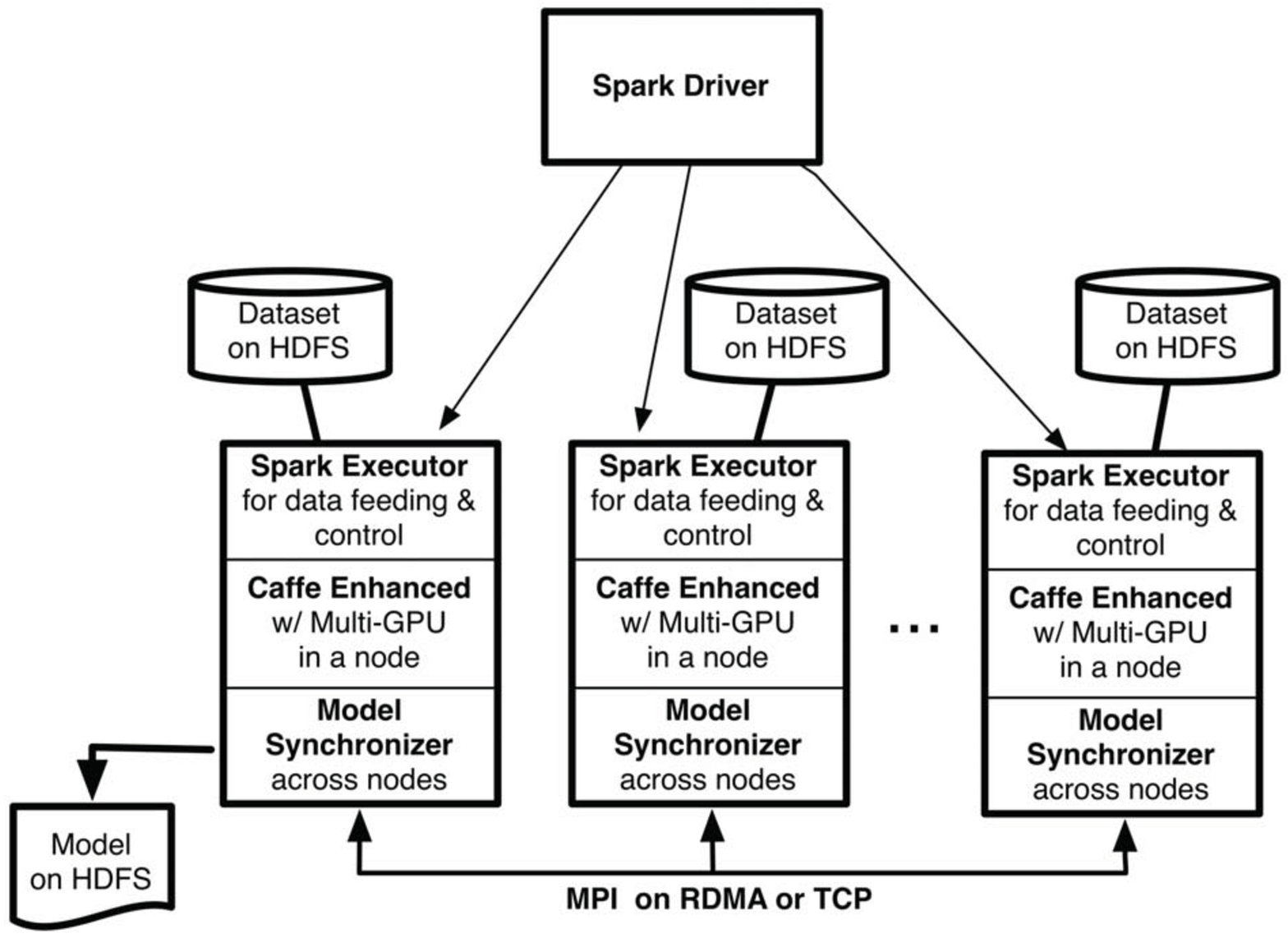}
		\vspace*{-0.05in}
		\caption{\small CaffeOnSpark Architecture.~\cite{CaffeOnSpark_Introduction}}
		\label{CaffeOnSpark_architecture}
	\end{center}
\end{figure}

The architecture of CaffeOnSpark is shown in Figure~\ref{CaffeOnSpark_architecture}. It supports the launch of Caffe engines on GPU or CPU devices within the Spark executor by invoking a JNI layer with fine-grain memory management. Moreover, it takes Spark+MPI architecture in order for CaffeOnSpark to achieve similar performance as dedicated deep learning clusters  by using MPI allreduce style interface via TCP/Ethernet or RDMA/Infiniband for the network communication across CaffeOnSpark executors.

\emph{2). Deeplearning4j/dl4j-spark-ml.} Deeplearning4j~\cite{deeplearning4j} is the first commercial-grade, open-source, distributed deep learning library written for Java and Scala, and a computing framework with the support and implementation of many deep learning algorithms, including restricted Boltzmann machine, deep belief net, deep autoencoder, stacked denoising autoencoder and recursive neural tensor network, word2vec, doc2vec and GloVe. It integrates with Spark via a Spark package called \emph{dl4j-spark-ml}~\cite{dl4j-spark-ml}, which provides a set of Spark components including DataFrame Readers for MNIST, Labeled Faces in the Wild (LFW) and IRIS, and pipeline components for NeuralNetworkClassification and NeuralNetworkReconstruction. It supports heterogeneous architecture by using Spark CPU to drive GPU coprocessors in a distributed context.

\emph{3). SparkNet.} SparkNet~\cite{moritz2015sparknet,SparkNet} is an open-source, distributed system for training deep network in Spark released by the AMPLab at U.C. Berkley in Nov 2015.
It is built on top of Spark and Caffe, where Spark is responsible for distributed data processing and the core learning process is delegated to the Caffe framework. SparkNet provides an interface for reading data from Spark RDDs and a compatible interface to the Caffe. It achieves a good scalability and tolerance of high-latency communication by using a simple palatalization scheme for stochastic gradient descent.  It also allows Spark users to construct deep networks using existing deep learning libraries or systems, such as TensorFlow~\cite{abadi2016tensorflow} or Torch as a backend, instead of building a new deep learning library in Java or Scala. Such a new integrated model of combining existing model training frameworks with existing batch frameworks is beneficial in practice. For example, machine learning often involves a set of pipeline tasks such as data retrieving, cleaning and processing before model training as well as model deployment and model prediction after training. All of these can be well handled with the existing data-processing pipelines in today's distributed computational environments such as Spark. Moreover, the integrated model of SparkNet can inherit the in-memory computation from Spark that allows data to be cached in memory from start to complete for fast computation, instead of writing to disk between operations as a segmented approach does. It also allows machining learning algorithm easily to pipeline with Spark's other components such as Spark SQL and GraphX.

Moreover, there are some other Spark-based deep learning libraries and frameworks, including OpenDL~\cite{OpenDL}, DeepDist~\cite{DeepDist}, dllib~\cite{dllib} , MMLSpark~\cite{MMLSpark}, and DeepSpark~\cite{kim2016deepspark}. OpenDL~\cite{OpenDL} is a deep learning training library based on Spark by applying the similar idea used by DistBelief~\cite{dean2012}. It executes the distributed training by splitting the training data into different data shards and synchronizes the replicate model using a centralized parameter server. DeepDist~\cite{DeepDist} accelerates model training by providing asynchronous stochastic gradient descent for data stored on HDFS / Spark. dllib~\cite{dllib} is a distributed deep learning framework based on Apache Spark. It provides a simple and easy-to-use interface for users to write and run deep learning algorithms on spark. MMLSpark~\cite{MMLSpark} provides a number of deep learning tools for Apache Spark, including seamless integration of Spark Machine Learning pipelines with Microsoft Cognitive Toolkit (CNTK) and OpenCV, enabling users to quickly create powerful, highly-scalable predictive and analytical models for large  and text datasets. DeepSpark~\cite{kim2016deepspark} is an alternative deep learning framework similar to SparkNet. It integrates three components including Spark, asynchronous parameter updates, and GPU-based Caffe seamlessly for enhanced large-scale data processing pipeline and accelerated DNN training.

\subsection{Spark Applications}
\label{Spark_Applications}

As an efficient data processing system, Spark has been widely used in many application domains, including Genomics, Medicine$\&$Healthcare, Finance, and Astronomy, etc.

\subsubsection{Genomics}
The method of the efficient score statistic is used extensively to conduct inference for high throughput genomic data due to its computational efficiency and ability to accommodate simple and complex phenotypes. To address the resulting computational challenge for resampling based inference, what is needed is a scalable and distributed computing approach. A cloud computing platform is suitable as it allows researchers to conduct data analyses at moderate costs, participating in the absence of access to a large computer infrastructure. SparkScore~\cite{7529900} is a set of distributed computational algorithms implemented in Apache Spark, to leverage the embarrassingly parallel nature of genomic resampling inference on the basis of the efficient score statistics. This computational approach harnesses the fault-tolerant features of Spark and can be readily extended to analysis of DNA and RNA sequencing data, including expression quantitative trait loci (eQTL) and phenotype association studies. Experiments conducted with Amazon's Elastic MapReduce (EMR) on synthetic data sets demonstrate the efficiency and scalability of SparkScore, including high-volume resampling of very large data sets.
To study the utility of Apache Spark in the genomic context, SparkSeq~\cite{wiewiorka2014sparkseq} was created. SparkSeq performs in-memory computations on the Cloud via Apache Spark. It covers operations on Binary Alignment/Map (BAM) and Sequence Alignment/Map (SAM) files, and it supports filtering of reads summarizing genomic features and basic statistical analyses operations. SparkSeq is a general-purpose tool for RNA and DNA sequencing analyses, tuned for processing in the cloud big alignment data with nucleotide precision. SparkSeq opens up the possibility of customized ad hoc secondary analyses and iterative machine learning algorithms.

\subsubsection{Medicine \& Healthcare}
In a modern society with great pressure, more and more people trapped in health issues. In order to reduce the cost of medical treatments, many organizations were devoted to adopting big data analytics into practice so as to avoid cost. Large amount of healthcare data is produced in  healthcare industry but the utilization of those data is low without processing this data interactively in real-time~\cite{Archenaa2016Interactive}. But now it is possible to process real time healthcare data because spark supports automated analytics through iterative processing on large data set. But in some circumstances the quality of data is poor, which brings a big problem. A spark-based approach to data processing and probabilistic record linkage is presented in order to produce very accurate data marts~\cite{Barreto2015A}. This approach is specifically on supporting the assessment of data quality, pre-processing, and linkage of databases provided by the Ministry of Health and the Ministry of Social Development and Hunger Alleviation.

%
%
%
\subsubsection{Finance}
Big data analytic technique is an effective way to provide good financial services for users in financial domain. For stock market, to have an accurate prediction and decision on the market trend, there are many factors such as politics and social events needed to be considered. Mohamed et al. \cite{10.1007/978-3-319-74690-6_66} propose a real-time prediction model of stock market trends by analyzing big data of news, tweets, and historical price with Apache Spark. The model supports the offline mode that works on historical data, and real-time mode that works on real-time data during the stock market session. Li et al.~\cite{Spark_in_Finance_Quantitative_Investing} builds a quantitative investing tool based on Spark that can be used for macro timing and portifolio rebalancing in the market.

To protect user's account during the digital payment and online transactions, fraud detection is a very important issue in financial service. Rajeshwari et al.\cite{7912039} study the credit card fraud detection. It takes Spark streaming data processing to provide real-time fraud detection based on Hidden Markov Model (HMM) during the credit card transaction by analyzing its log data and new generated data. Carcillo et al.~\cite{CARCILLO2018182} propose a realistic and scalable fraud detection system called Real-time Fraud Finder (SCARFF), which integrates Big Data software (Kafka, Spark and Cassandra) with a machine learning approach that deals with class imbalance, nonstationarity and verification latency.

Moreover, there are some other financial applications such as financial risk analysis~\cite{Financial_Risk_Estimation}, financial trading~\cite{Dutta_2015}, etc.


%
%
%

\subsubsection{Astronomy}
Considering the technological advancement of telescopes and the number of ongoing sky survey projects, it is safe to say that astronomical research is moving into the Big Data era. The sky surveys deliver huge datasets that can be used for different scientific studies simultaneously. Kira~\cite{zhang2015scientific}, a flexible and distributed astronomy  processing toolkit using Apache Spark, is proposed to implement a Source Extractor application for astronomy s. The extraction accuracy can be improved by running multiple iterations of source extraction.

\begin{figure}[h]
	\begin{center}
		\includegraphics
		[width=0.4\textwidth]{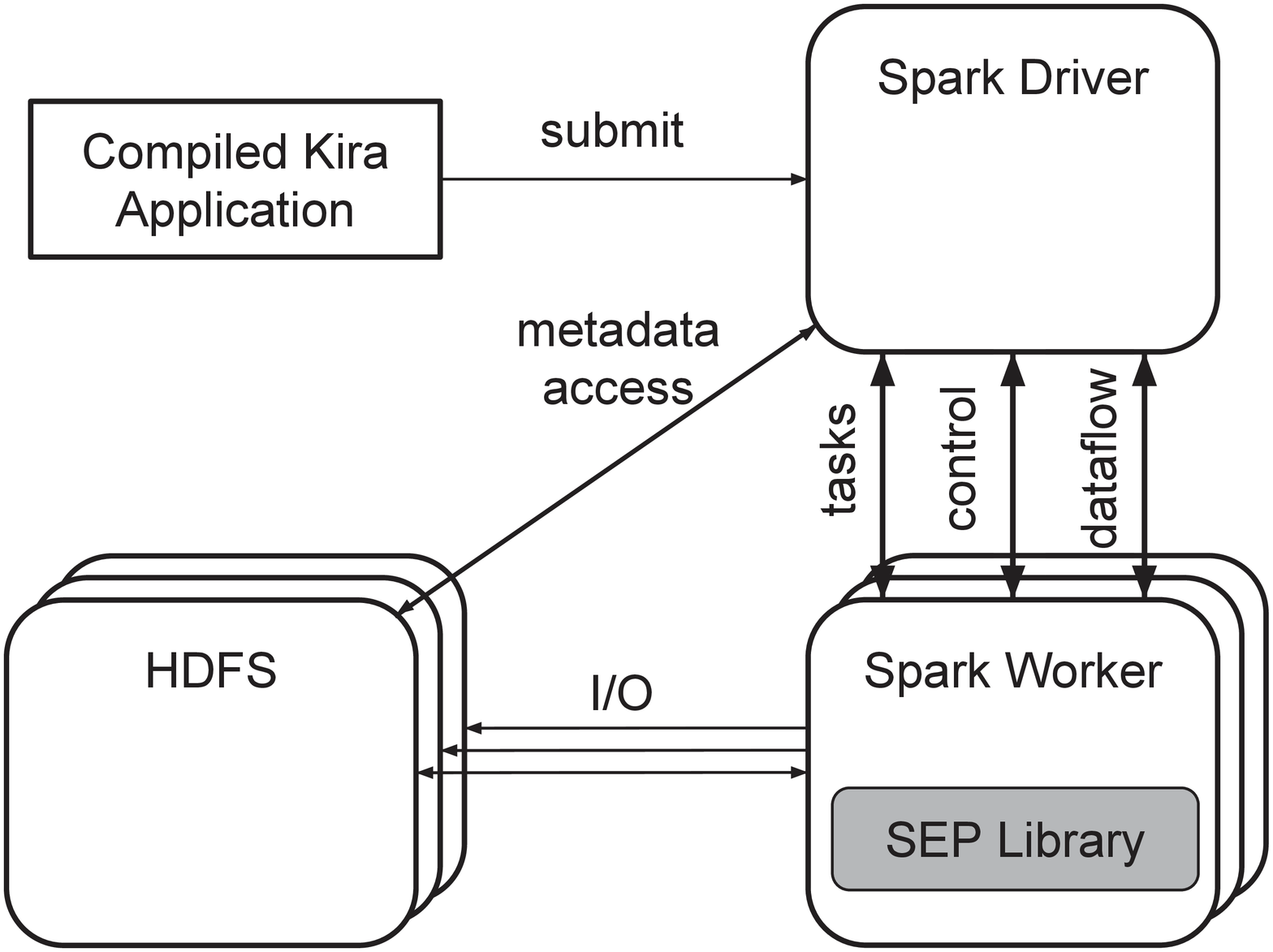}
		\vspace*{-0.05in}
		\caption{\small The Overview of Kira Architecture.~\cite{zhang2016kira}}
		\label{Kira_Architecture}
	\end{center}
\end{figure}

Figure 1 shows the architecture of Kira and inter-component interactions. Kira runs on top of Spark, which supports a single driver and multiple workers and the SEP library is deployed to all worker nodes~\cite{zhang2016kira}. Kira reimplements the Source Extractor algorithm from scratch and connects existing programs as monolithic pieces. The approach is exposed a programmable library and allows users to reuse the legacy code without sacrificing control-flow flexibility. The Kira SE implementation demonstrates linear scalability with the dataset and cluster size.

The huge volume and rapid growth of dataset in scientific computing such as Astronomy demand for a fast and scalable data processing system. Leveraging a big data platform such as Spark would enable scientists to benefit from the rapid pace of innovation and large range of systems that are being driven by widespread interest in big data analytics.

\section{Challenges and Open Issues}
\label{Challenges_and_Open_Issues}

In this section, we discuss the research challenges and opportunities for Spark ecosystem.

%
\emph{Memory Resource Management.}
As an in-memory processing platform built with Scala, Spark's performance is sensitive to its memory configuration and usage of JVMs. The memory resource is divided into two parts. One is for RDD caching. The other is used for tasks' working memory to store objects created during the task execution. The proper configuration of such memory allocation is non-trivial for performance improvement. Moreover, the overhead of JVM garbage collection (GC) can be a problem when there are a large number of ``churn'' for cached RDDs, or due to serious interference between the cached RDDs and tasks' working memory. For this, Maas \emph{et al}~\cite{maas2016taurus} have a detailed study for GC's impact on Spark in distributed environment. The proper tuning of GC thus plays an important role in performance optimization. Currently, it is still at early stage and there are not good solutions for Spark. It opens an important issue on the memory resource management and GC tuning for Spark. Regarding this, recently, Spark community starts a new project for Spark called Tungsten~\cite{tungsten} that places Spark's memory management as its first concern.


\emph{New Emerging Processor Support.}
In addition to GPU and FPGA, the recent advancement on computing hardware make some new processors emerged, such as APU~\cite{foley2011amd} and TPU~\cite{Jouppi:2017:IPA:3079856.3080246}, etc. These can bring new opportunities to enhance the performance of Spark system. For example, APU is a coupled CPU-GPU device that integrates the CPU and the GPU into a single chip and allows the CPU and the GPU to communicate with each other through the shared physical memory by featuring shared memory space between them~\cite{foley2011amd}. It can improve the performance of existing discrete CPU-GPU architecture where CPU and GPU communicate via PCI-e bus. TPU is a domain-specific processor for deep neural network. It can give us a chance to speedup Spark for deep learning applications by migrating Spark to TPU platform.

\emph{Heterogenous Accelerators Support.} Besides emerging processors, it could be possible in practice that a Spark computing system consists of a number of diverse processors such as CPU, GPU, FPGA and MIC as illustrated in Spark ecosystem of Figure~\ref{Spark_Support}. Rather than supporting a single processor only, it is crucial to have a upgraded Spark that can utilize all of the computing devices simultaneously for maximum performance. Due to the fact that different accelerators are based on different programming models (e.g., CUDA for GPU, OpenCL for FPGA), it open us a new challenge on how to support such different types of accelerators for Spark at the same time.



\emph{RDD Operation and Sharing.}
There are several open issues for current Spark's RDD. First, it allows only coarse-grained operations (i.e., one operation for all data) on RDDs, whereas the fine-grained operations (e.g., partial read) are supported. One work is to design some fine-grained operations on partial data of RDD. Second, current RDDs are immutable. Instead of modifying on existing RDD, any update operation would generate new RDD, some data of which can be redundant and thus results in a wast of storage resource. Third, for a RDD, its data partitions can be skewed, i.e., there are many small partitions coupled with a few number of large-size partitions. Moreover, a Spark task computation generally involves a series of pipelined RDDs. Thus, the skewed RDD partitions can easily incur the chained unbalanced problem for tasks, which causes some workers much busier than others. Fourth, Spark itself does not support RDD sharing across applications. For some applications that have the same input data or redundant task computation, enabling RDD sharing can be an effective approach to improve the performance of the whole applications.

\emph{Failure Recovery.}
%
%
In contrast to MapReduce that provides fault tolerance through replication or checkpoint, Spark achieves failure recovery via lineage re-computation, which is much more cost efficient since it saves the costs due to data replication across the network and disk storage. The lineage information (e.g., input data, computing function) for each RDD partition is recorded. Any lost data of RDDs can be recovered through re-computation based on its lineage information. However, there is a key assumption that all RDD lineage information is kept and always available, and the driver does not fail. It means that Spark is not $100\%$ fault tolerance without overcoming this assumption. It thus remains us an open issue on how to enhance fault tolerance for Spark.

%
%

%
%

\section{Conclusion}
\label{Conclusion}
Spark has gained significant interests and contributions both from industry and academia because of its simplicity, generality, fault tolerance, and high performance. However, there is a lack of work to summarize and classify them comprehensively. In view of this, it motives us to investigate the related work on Spark. We first overview the Spark framework, and present the pros and cons of Spark. We then provide a comprehensive review of the current status of Spark studies and related work in the literature that aim at improving and enhancing the Spark framework, and give the open issues and challenges regarding the current Spark finally. In summary, we hopefully expect to see that this work can be a useful resource for users who are interested in Spark and want to have further study on Spark.

\bibliographystyle{plain}
\bibliography{./references}

\begin{thebibliography}{100}

\bibitem{flink}
Apache flink, url: https://flink.apache.org/.

\bibitem{spark2}
Apache spark as a compiler: Joining a billion rows per second on a laptop.
\newblock In {\em
  https://databricks.com/blog/2016/05/23/apache-spark-as-a-compiler-joining-a-billion-rows-per-second-on-a-laptop.html}.

\bibitem{Decision}
Decision cep url: http//github.com/stratio/decision.

\bibitem{tungsten}
Project tungsten: Bringing apache spark closer to bare metal.
\newblock In {\em
  https://databricks.com/blog/2015/04/28/project-tungsten-bringing-spark-closer-to-bare-metal.html}.

\bibitem{Sparkcep}
Spark cep url: https://github.com/samsung/spark-cep.

\bibitem{streamDM}
Streamdm, url: http: http//huawei-noah.github.io/streamdm/.

\bibitem{Financial_Risk_Estimation}
Estimating financial risk with apache spark.
\newblock In {\em
  https://blog.cloudera.com/blog/2014/07/estimating-financial-risk-with-apache-spark/},
  2014.

\bibitem{Hive_on_Spark_analysis}
Shark, spark sql, hive on spark, and the future of sql on apache spark.
\newblock In {\em
  https://databricks.com/blog/2014/07/01/shark-spark-sql-hive-on-spark-and-the-future-of-sql-on-spark.html},
  2014.

\bibitem{HBase}
Apache hbase.
\newblock In {\em http://hbase.apache.org/}, 2015.

\bibitem{Knox}
Apache knox gateway.
\newblock In {\em http://hortonworks.com/hadoop/knox-gateway/}, 2015.

\bibitem{Apache_Ranger}
Apache ranger.
\newblock In {\em http://hortonworks.com/hadoop/ranger/}, 2015.

\bibitem{Spark_Security}
Apache security.
\newblock In {\em https://spark.apache.org/docs/latest/security.html}, 2015.

\bibitem{Spark_homepage}
Apache spark.
\newblock In {\em https://spark.apache.org/}, 2015.

\bibitem{Storm}
Apache storm.
\newblock In {\em https://storm.apache.org/}, 2015.

\bibitem{DeepDist}
Deepdist: Lightning-fast deep learning on spark via parallel stochastic
  gradient updates.
\newblock In {\em http://deepdist.com/}, 2015.

\bibitem{Sentry}
Introducing sentry.
\newblock In {\em
  http://www.cloudera.com/content/cloudera/en/\\campaign/introducing-sentry.html},
  2015.

\bibitem{MLlib}
Machine learning library (mllib) guide.
\newblock In {\em MLib: https://spark.apache.org/docs/latest/mllib-guide.html},
  2015.

\bibitem{OpenDL}
Opendl: The deep learning training framework on spark.
\newblock In {\em https://github.com/guoding83128/OpenDL/}, 2015.

\bibitem{Alluxio}
Alluxio, formerly known as tachyon, is a memory speed virtual distributed
  storage system.
\newblock In {\em http://www.alluxio.org/}, 2016.

\bibitem{Amazon_DynamoDB}
Amazon dynamodb.
\newblock In {\em https://en.wikipedia.org/wiki/Amazon\_DynamoDB}, 2016.

\bibitem{Amazon_S3}
Amazon s3.
\newblock In {\em https://en.wikipedia.org/wiki/Amazon\_S3}, 2016.

\bibitem{Apache_Cassandra}
Apache cassandra.
\newblock In {\em https://en.wikipedia.org/wiki/Apache\_Cassandra}, 2016.

\bibitem{Apache_Hive}
Apache hive.
\newblock In {\em https://github.com/apache/hive}, 2016.

\bibitem{Apache_Pig}
Apache pig.
\newblock In {\em https://pig.apache.org/}, 2016.

\bibitem{CaffeOnSpark}
Caffeonspark.
\newblock In {\em https://github.com/yahoo/CaffeOnSpark}, 2016.

\bibitem{CaffeOnSpark_Introduction}
Caffeonspark open sourced for distributed deep learning on big data clusters.
\newblock In {\em
  http://yahoohadoop.tumblr.com/post/139916563586/caffeonspark-open-sourced-for-distributed-deep},
  2016.

\bibitem{Cloud_storage}
Cloud storage.
\newblock In {\em https://en.wikipedia.org/wiki/Cloud\_storage}, 2016.

\bibitem{Distributed_hash_table}
Distributed hash table.
\newblock In {\em https://en.wikipedia.org/wiki/Distributed\_hash\_table},
  2016.

\bibitem{SparkNet}
Distributed neural networks for spark.
\newblock In {\em https://github.com/amplab/SparkNet}, 2016.

\bibitem{DynamoDB_Spark}
Dynamodb data source for apache spark.
\newblock In {\em https://github.com/traviscrawford/spark-dynamodb}, 2016.

\bibitem{Synapse}
Encode-dream in-vivo transcription factor binding site prediction challenge.
\newblock In {\em https://www.synapse.org/\#!Synapse:syn6131484}, 2016.

\bibitem{Freeman_Lab}
Freeman lab.
\newblock In {\em https://www.janelia.org/lab/freeman-lab}, 2016.

\bibitem{H2O}
H2o.
\newblock In {\em https://github.com/h2oai/h2o-3}, 2016.

\bibitem{H2O.ai}
H2o.ai.
\newblock In {\em http://www.h2o.ai/}, 2016.

\bibitem{Azure_Storage}
Introduction to microsoft azure storage.
\newblock In {\em
  https://azure.microsoft.com/en-us/documentation/articles/storage-introduction/},
  2016.

\bibitem{Medium}
Medium.
\newblock In {\em https://medium.com/}, 2016.

\bibitem{deeplearning4j}
Open-source, distributed, deep-learning library for the jvm.
\newblock In {\em http://deeplearning4j.org/}, 2016.

\bibitem{PySpark_Cassandra}
Pyspark cassandra.
\newblock In {\em https://github.com/TargetHolding/pyspark-cassandra}, 2016.

\bibitem{R_programmingLanguage}
The r project for statistical computing.
\newblock In {\em https://www.r-project.org/}, 2016.

\bibitem{AmazonS3}
S3 support in apache hadoop.
\newblock In {\em http://wiki.apache.org/hadoop/AmazonS3}, 2016.

\bibitem{Scala_Language}
Scala language.
\newblock In {\em https://spark.apache.org/docs/latest/api/python/index.html},
  2016.

\bibitem{Spark_Cassandra_Connector}
Spark cassandra connector.
\newblock In {\em https://github.com/datastax/spark-cassandra-connector}, 2016.

\bibitem{spark_gpu_columnar_ibm}
Spark-gpu wiki.
\newblock In {\em https://github.com/kiszk/spark-gpu}, 2016.

\bibitem{Spark_HBase_Connector}
Spark-hbase connector.
\newblock In {\em https://github.com/nerdammer/spark-hbase-connector}, 2016.

\bibitem{Spark_in_Finance_Quantitative_Investing}
Spark-in-finance-quantitative-investing.
\newblock In {\em
  https://github.com/litaotao/Spark-in-Finance-Quantitative-Investing}, 2016.

\bibitem{spark-on-hbase}
spark-on-hbase.
\newblock In {\em https://github.com/michal-harish/spark-on-hbase}, 2016.

\bibitem{dl4j-spark-ml}
Spark package - dl4j-spark-ml.
\newblock In {\em https://github.com/deeplearning4j/dl4j-spark-ml}, 2016.

\bibitem{PySpark}
Spark python api.
\newblock In {\em http://spark.apache.org/docs/latest/api/python/index.html},
  2016.

\bibitem{Python_Spark}
Spark python api docs.
\newblock In {\em http://www.scala-lang.org/}, 2016.

\bibitem{spark_S3}
spark-s3.
\newblock In {\em https://github.com/knoldus/spark-s3}, 2016.

\bibitem{Spark-SQL-on-HBase}
Spark-sql-on-hbase.
\newblock In {\em https://github.com/Huawei-Spark/Spark-SQL-on-HBase}, 2016.

\bibitem{sparkling-water}
Sparkling water.
\newblock In {\em https://github.com/h2oai/sparkling-water}, 2016.

\bibitem{SparkR}
Sparkr (r on spark).
\newblock In {\em https://spark.apache.org/docs/latest/sparkr.html}, 2016.

\bibitem{Spork}
Spork: Pig on apache spark.
\newblock In {\em https://github.com/sigmoidanalytics/spork}, 2016.

\bibitem{Thunder}
Thunder: Large-scale analysis of neural data.
\newblock In {\em http://thunder-project.org/}, 2016.

\bibitem{ADAM}
Adam.
\newblock In {\em http://adam.readthedocs.io/en/adam-parent/$\_2.11-0.23.0$/},
  2017.

\bibitem{dllib}
dllib.
\newblock In {\em https://github.com/Lewuathe/dllib}, 2017.

\bibitem{KeystoneML}
Keystoneml api docs.
\newblock In {\em http://keystone-ml.org/}, 2017.

\bibitem{SSD_caching_DataBricks}
Databricks cache boosts apache spark performance-why nvme ssds improve caching
  performance by 10x.
\newblock In {\em
  https://databricks.com/blog/2018/01/09/databricks-cache-boosts-apache-spark-performance.html},
  2018.

\bibitem{MMLSpark}
Mmlspark: Microsoft machine learning for apache spark.
\newblock In {\em https://github.com/Azure/mmlspark}, 2018.

\bibitem{Davidson}
Davidson Aaron and Or~Andrew.
\newblock Optimizing shuffle performance in spark.
\newblock In {\em University of California, Berkeley - Department of Electrical
  Engineering and Computer Sciences}, 2013.

\bibitem{abadi2016tensorflow}
Mart{\i}n Abadi, Ashish Agarwal, Paul Barham, Eugene Brevdo, Zhifeng Chen,
  Craig Citro, Greg~S Corrado, Andy Davis, Jeffrey Dean, Matthieu Devin, et~al.
\newblock Tensorflow: Large-scale machine learning on heterogeneous distributed
  systems.
\newblock {\em arXiv preprint arXiv:1603.04467}, 2016.

\bibitem{Succinct}
Rachit Agarwal, Anurag Khandelwal, and Ion Stoica.
\newblock Succinct: Enabling queries on compressed data.
\newblock In {\em 12th USENIX Symposium on Networked Systems Design and
  Implementation (NSDI 15)}, pages 337--350, Oakland, CA, May 2015. USENIX
  Association.

\bibitem{Agarwal2014Knowing}
Sameer Agarwal, Henry Milner, Ariel Kleiner, Ameet Talwalkar, Michael Jordan,
  Samuel Madden, Barzan Mozafari, and Ion Stoica.
\newblock Knowing when you're wrong: Building fast and reliable approximate
  query processing systems.
\newblock {\em Association for Computing Machinery}, pages 481--492, 2014.

\bibitem{BlinkDB}
Sameer Agarwal, Barzan Mozafari, Aurojit Panda, Henry Milner, Samuel Madden,
  and Ion Stoica.
\newblock Blinkdb: Queries with bounded errors and bounded response times on
  very large data.
\newblock In {\em Proceedings of the 8th ACM European Conference on Computer
  Systems}, EuroSys '13, pages 29--42, New York, NY, USA, 2013. ACM.

\bibitem{Archenaa2016Interactive}
J.~Archenaa and E.~A.~Mary Anita.
\newblock {\em Interactive Big Data Management in Healthcare Using Spark}.
\newblock Springer International Publishing, 2016.

\bibitem{Badam:2011:SHS:1972457.1972479}
Anirudh Badam and Vivek~S. Pai.
\newblock Ssdalloc: Hybrid ssd/ram memory management made easy.
\newblock In {\em Proceedings of the 8th USENIX Conference on Networked Systems
  Design and Implementation}, NSDI'11, pages 211--224, Berkeley, CA, USA, 2011.
  USENIX Association.

\bibitem{7529900}
A.~Bahmani, A.~B. Sibley, M.~Parsian, K.~Owzar, and F.~Mueller.
\newblock Sparkscore: Leveraging apache spark for distributed genomic
  inference.
\newblock In {\em 2016 IEEE International Parallel and Distributed Processing
  Symposium Workshops (IPDPSW)}, pages 435--442, May 2016.

\bibitem{Barreto2015A}
Marcos Barreto, Robespierre Pita, Clicia Pinto, Malu Silva, Pedro Melo, and
  Davide Rasella.
\newblock A spark-based workflow for probabilistic record linkage of healthcare
  data.
\newblock In {\em The Workshop on Algorithms \& Systems for Mapreduce \&
  Beyond}, 2015.

\bibitem{7395869}
A.~Bifet, S.~Maniu, J.~Qian, G.~Tian, C.~He, and W.~Fan.
\newblock Streamdm: Advanced data mining in spark streaming.
\newblock In {\em 2015 IEEE International Conference on Data Mining Workshop
  (ICDMW)}, pages 1608--1611, Nov 2015.

\bibitem{Bose:2005:LRS:1057977.1057978}
Rajendra Bose and James Frew.
\newblock Lineage retrieval for scientific data processing: A survey.
\newblock {\em ACM Comput. Surv.}, 37(1):1--28, March 2005.

\bibitem{foley2011amd}
Alexander Branover, Denis Foley, and Maurice Steinman.
\newblock Amd fusion apu: Llano.
\newblock {\em IEEE Micro}, 32(2):28--37, March 2012.

\bibitem{CARCILLO2018182}
Fabrizio Carcillo, Andrea~Dal Pozzolo, Yann-Aël~Le Borgne, Olivier Caelen,
  Yannis Mazzer, and Gianluca Bontempi.
\newblock Scarff: A scalable framework for streaming credit card fraud
  detection with spark.
\newblock {\em Information Fusion}, 41:182 -- 194, 2018.

\bibitem{Redis}
Josiah~L. Carlson.
\newblock {\em Redis in Action}.
\newblock Manning Publications Co., Greenwich, CT, USA, 2013.

\bibitem{Chang:2006:BDS:1267308.1267323}
Fay Chang, Jeffrey Dean, Sanjay Ghemawat, Wilson~C. Hsieh, Deborah~A. Wallach,
  Mike Burrows, Tushar Chandra, Andrew Fikes, and Robert~E. Gruber.
\newblock Bigtable: A distributed storage system for structured data.
\newblock In {\em Proceedings of the 7th USENIX Symposium on Operating Systems
  Design and Implementation - Volume 7}, OSDI '06, pages 15--15, Berkeley, CA,
  USA, 2006. USENIX Association.

\bibitem{spark_fpga_hotcloud16}
Yu-Ting Chen, Jason Cong, Zhenman Fang, Jie Lei, and Peng Wei.
\newblock When spark meets fpgas: A case study for next-generation dna
  sequencing acceleration.
\newblock In {\em 8th USENIX Workshop on Hot Topics in Cloud Computing
  (HotCloud 16)}, Denver, CO, June 2016. USENIX Association.

\bibitem{8310322}
W.~Cheong, C.~Yoon, S.~Woo, K.~Han, D.~Kim, C.~Lee, Y.~Choi, S.~Kim, D.~Kang,
  G.~Yu, J.~Kim, J.~Park, K.~W. Song, K.~T. Park, S.~Cho, H.~Oh, D.~D.~G. Lee,
  J.~H. Choi, and J.~Jeong.
\newblock A flash memory controller for 15us ultra-low-latency ssd using
  high-speed 3d nand flash with 3us read time.
\newblock In {\em 2018 IEEE International Solid - State Circuits Conference -
  (ISSCC)}, pages 338--340, Feb 2018.

\bibitem{Vispark_isldav15}
W.~Choi and W.~K. Jeong.
\newblock Vispark: Gpu-accelerated distributed visual computing using spark.
\newblock In {\em Large Data Analysis and Visualization (LDAV), 2015 IEEE 5th
  Symposium on}, pages 125--126, Oct 2015.

\bibitem{fpga_dacecenter_dac16}
Jason Cong, Muhuan Huang, Di~Wu, and Cody~Hao Yu.
\newblock Invited - heterogeneous datacenters: Options and opportunities.
\newblock In {\em Proceedings of the 53rd Annual Design Automation Conference},
  DAC '16, pages 16:1--16:6, New York, NY, USA, 2016. ACM.

\bibitem{Crankshaw2014The}
Daniel Crankshaw, Peter Bailis, Joseph~E. Gonzalez, Haoyuan Li, Zhao Zhang,
  Michael~J. Franklin, Ali Ghodsi, and Michael~I. Jordan.
\newblock The missing piece in complex analytics: Low latency, scalable model
  management and serving with velox.
\newblock {\em European Journal of Obstetrics \& Gynecology \& Reproductive
  Biology}, 185:181--182, 2014.

\bibitem{das2014adaptive}
Tathagata Das, Yuan Zhong, Ion Stoica, and Scott Shenker.
\newblock Adaptive stream processing using dynamic batch sizing.
\newblock In {\em Proceedings of the ACM Symposium on Cloud Computing}, pages
  1--13. ACM, 2014.

\bibitem{dean2012}
Jeffrey Dean, Greg Corrado, Rajat Monga, Kai Chen, Matthieu Devin, Quoc~V. Le,
  Mark~Z. Mao, Marc'Aurelio Ranzato, Andrew~W. Senior, Paul~A. Tucker, Ke~Yang,
  and Andrew~Y. Ng.
\newblock Large scale distributed deep networks.
\newblock In {\em NIPS'12}, pages 1232--1240, 2012.

\bibitem{Dean}
Jeffrey Dean and Sanjay Ghemawat.
\newblock Mapreduce: Simplified data processing on large clusters.
\newblock In {\em Proceedings of the 6th Conference on Symposium on Opearting
  Systems Design \& Implementation - Volume 6}, OSDI'04, pages 10--10,
  Berkeley, CA, USA, 2004. USENIX Association.

\bibitem{DeCandia:2007:DAH:1323293.1294281}
Giuseppe DeCandia, Deniz Hastorun, Madan Jampani, Gunavardhan Kakulapati,
  Avinash Lakshman, Alex Pilchin, Swaminathan Sivasubramanian, Peter Vosshall,
  and Werner Vogels.
\newblock Dynamo: Amazon's highly available key-value store.
\newblock {\em SIGOPS Oper. Syst. Rev.}, 41(6):205--220, October 2007.

\bibitem{Dutta_2015}
Kamalika Dutta and Manasi Jayapal.
\newblock Big data analytics for real time systems, 02 2015.

\bibitem{Engle:2012:SFD:2213836.2213934}
Cliff Engle, Antonio Lupher, Reynold Xin, Matei Zaharia, Michael~J. Franklin,
  Scott Shenker, and Ion Stoica.
\newblock Shark: Fast data analysis using coarse-grained distributed memory.
\newblock In {\em Proceedings of the 2012 ACM SIGMOD International Conference
  on Management of Data}, SIGMOD '12, pages 689--692, New York, NY, USA, 2012.
  ACM.

\bibitem{cunningham2014analyzing}
Jeremy Freeman, Nikita Vladimirov, Takashi Kawashima, Yu~Mu, Nicholas~J
  Sofroniew, Davis~V Bennett, Joshua Rosen, Chao-Tsung Yang, Loren~L Looger,
  and Misha~B Ahrens.
\newblock Mapping brain activity at scale with cluster computing.
\newblock {\em Nature methods}, 11(9):941--950, 2014.

\bibitem{Gonzalez:2012:PDG:2387880.2387883}
Joseph~E. Gonzalez, Yucheng Low, Haijie Gu, Danny Bickson, and Carlos Guestrin.
\newblock Powergraph: Distributed graph-parallel computation on natural graphs.
\newblock In {\em Proceedings of the 10th USENIX Conference on Operating
  Systems Design and Implementation}, OSDI'12, pages 17--30, Berkeley, CA, USA,
  2012. USENIX Association.

\bibitem{GraphX}
Joseph~E. Gonzalez, Reynold~S. Xin, Ankur Dave, Daniel Crankshaw, Michael~J.
  Franklin, and Ion Stoica.
\newblock Graphx: Graph processing in a distributed dataflow framework.
\newblock In {\em Proceedings of the 11th USENIX Conference on Operating
  Systems Design and Implementation}, OSDI'14, pages 599--613, Berkeley, CA,
  USA, 2014. USENIX Association.

\bibitem{Gulzar2016BigDebug}
Muhammad~Ali Gulzar, Matteo Interlandi, Seunghyun Yoo, Sai~Deep Tetali, Tyson
  Condie, Todd Millstein, and Miryung Kim.
\newblock Bigdebug: Debugging primitives for interactive big data processing in
  spark.
\newblock In {\em Ieee/acm International Conference on Software Engineering},
  pages 784--795, 2016.

\bibitem{Gunda:2010:NAM:1924943.1924949}
Pradeep~Kumar Gunda, Lenin Ravindranath, Chandramohan~A. Thekkath, Yuan Yu, and
  Li~Zhuang.
\newblock Nectar: Automatic management of data and computation in datacenters.
\newblock In {\em Proceedings of the 9th USENIX Conference on Operating Systems
  Design and Implementation}, OSDI'10, pages 75--88, Berkeley, CA, USA, 2010.
  USENIX Association.

\bibitem{Mesos}
Benjamin Hindman, Andy Konwinski, Matei Zaharia, Ali Ghodsi, Anthony~D. Joseph,
  Randy Katz, Scott Shenker, and Ion Stoica.
\newblock Mesos: A platform for fine-grained resource sharing in the data
  center.
\newblock In {\em Proceedings of the 8th USENIX Conference on Networked Systems
  Design and Implementation}, NSDI'11, pages 295--308, Berkeley, CA, USA, 2011.
  USENIX Association.

\bibitem{8107572}
Z.~Hu, B.~Li, and J.~Luo.
\newblock Time- and cost- efficient task scheduling across geo-distributed data
  centers.
\newblock {\em IEEE Transactions on Parallel and Distributed Systems},
  29(3):705--718, March 2018.

\bibitem{Interlandi2017Adding}
Matteo Interlandi, Ari Ekmekji, Kshitij Shah, Muhammad~Ali Gulzar, Sai~Deep
  Tetali, Miryung Kim, Todd Millstein, and Tyson Condie.
\newblock Adding data provenance support to apache spark.
\newblock {\em Vldb Journal}, (4):1--21, 2017.

\bibitem{Interlandi2015Titian}
Matteo Interlandi, Kshitij Shah, Sai~Deep Tetali, Muhammad~Ali Gulzar,
  Seunghyun Yoo, Miryung Kim, Todd Millstein, and Tyson Condie.
\newblock Titian: Data provenance support in spark.
\newblock {\em Proceedings of the Vldb Endowment}, 9(3):216--227, 2015.

\bibitem{Isard:2007:DDD:1272996.1273005}
Michael Isard, Mihai Budiu, Yuan Yu, Andrew Birrell, and Dennis Fetterly.
\newblock Dryad: Distributed data-parallel programs from sequential building
  blocks.
\newblock In {\em Proceedings of the 2Nd ACM SIGOPS/EuroSys European Conference
  on Computer Systems 2007}, EuroSys '07, pages 59--72, New York, NY, USA,
  2007. ACM.

\bibitem{jia2014caffe}
Yangqing Jia, Evan Shelhamer, Jeff Donahue, Sergey Karayev, Jonathan Long, Ross
  Girshick, Sergio Guadarrama, and Trevor Darrell.
\newblock Caffe: Convolutional architecture for fast feature embedding.
\newblock {\em arXiv preprint arXiv:1408.5093}, 2014.

\bibitem{2016AGUFMSH34A02J}
E.~{Jonas}, V.~{Shankar}, M.~{Bobra}, and B.~{Recht}.
\newblock {Flare Prediction Using Photospheric and Coronal Image Data}.
\newblock {\em AGU Fall Meeting Abstracts}, December 2016.

\bibitem{Jouppi:2017:IPA:3079856.3080246}
Norman~P. Jouppi, Cliff Young, Nishant Patil, David Patterson, Gaurav Agrawal,
  Raminder Bajwa, Sarah Bates, Suresh Bhatia, Nan Boden, Al~Borchers, Rick
  Boyle, Pierre-luc Cantin, Clifford Chao, Chris Clark, Jeremy Coriell, Mike
  Daley, Matt Dau, Jeffrey Dean, Ben Gelb, Tara~Vazir Ghaemmaghami, Rajendra
  Gottipati, William Gulland, Robert Hagmann, C.~Richard Ho, Doug Hogberg, John
  Hu, Robert Hundt, Dan Hurt, Julian Ibarz, Aaron Jaffey, Alek Jaworski,
  Alexander Kaplan, Harshit Khaitan, Daniel Killebrew, Andy Koch, Naveen Kumar,
  Steve Lacy, James Laudon, James Law, Diemthu Le, Chris Leary, Zhuyuan Liu,
  Kyle Lucke, Alan Lundin, Gordon MacKean, Adriana Maggiore, Maire Mahony,
  Kieran Miller, Rahul Nagarajan, Ravi Narayanaswami, Ray Ni, Kathy Nix, Thomas
  Norrie, Mark Omernick, Narayana Penukonda, Andy Phelps, Jonathan Ross, Matt
  Ross, Amir Salek, Emad Samadiani, Chris Severn, Gregory Sizikov, Matthew
  Snelham, Jed Souter, Dan Steinberg, Andy Swing, Mercedes Tan, Gregory
  Thorson, Bo~Tian, Horia Toma, Erick Tuttle, Vijay Vasudevan, Richard Walter,
  Walter Wang, Eric Wilcox, and Doe~Hyun Yoon.
\newblock In-datacenter performance analysis of a tensor processing unit.
\newblock In {\em Proceedings of the 44th Annual International Symposium on
  Computer Architecture}, ISCA '17, pages 1--12, New York, NY, USA, 2017. ACM.

\bibitem{kim2016deepspark}
Hanjoo Kim, Jaehong Park, Jaehee Jang, and Sungroh Yoon.
\newblock Deepspark: Spark-based deep learning supporting asynchronous updates
  and caffe compatibility.
\newblock {\em arXiv preprint arXiv:1602.08191}, 2016.

\bibitem{DBLP:journals/corr/abs-1708-05746}
Mijung Kim, Jun Li, Haris Volos, Manish Marwah, Alexander Ulanov, Kimberly
  Keeton, Joseph Tucek, Lucy Cherkasova, Le~Xu, and Pradeep Fernando.
\newblock Sparkle: Optimizing spark for large memory machines and analytics.
\newblock {\em CoRR}, abs/1708.05746, 2017.

\bibitem{Kleiner:2013:GBP:2487575.2487650}
Ariel Kleiner, Ameet Talwalkar, Sameer Agarwal, Ion Stoica, and Michael~I.
  Jordan.
\newblock A general bootstrap performance diagnostic.
\newblock In {\em Proceedings of the 19th ACM SIGKDD International Conference
  on Knowledge Discovery and Data Mining}, KDD '13, pages 419--427, New York,
  NY, USA, 2013. ACM.

\bibitem{kraska2013mlbase}
Tim Kraska, Ameet Talwalkar, John~C Duchi, Rean Griffith, Michael~J Franklin,
  and Michael~I Jordan.
\newblock Mlbase: A distributed machine-learning system.
\newblock In {\em CIDR}, volume~1, pages 2--1, 2013.

\bibitem{Krishnan:2016:IDA:2872427.2883026}
Dhanya~R. Krishnan, Do~Le Quoc, Pramod Bhatotia, Christof Fetzer, and Rodrigo
  Rodrigues.
\newblock Incapprox: A data analytics system for incremental approximate
  computing.
\newblock In {\em Proceedings of the 25th International Conference on World
  Wide Web}, WWW '16, pages 1133--1144, Republic and Canton of Geneva,
  Switzerland, 2016. International World Wide Web Conferences Steering
  Committee.

\bibitem{Cassandra}
Avinash Lakshman and Prashant Malik.
\newblock Cassandra: A decentralized structured storage system.
\newblock {\em SIGOPS Oper. Syst. Rev.}, 44(2):35--40, April 2010.

\bibitem{Lam:2012:MMP:2367502.2367520}
Wang Lam, Lu~Liu, Sts Prasad, Anand Rajaraman, Zoheb Vacheri, and AnHai Doan.
\newblock Muppet: Mapreduce-style processing of fast data.
\newblock {\em Proc. VLDB Endow.}, 5(12):1814--1825, August 2012.

\bibitem{2017arXiv170902946L}
D.~{Le Quoc}, R.~{Chen}, P.~{Bhatotia}, C.~{Fetze}, V.~{Hilt}, and T.~{Strufe}.
\newblock {Approximate Stream Analytics in Apache Flink and Apache Spark
  Streaming}.
\newblock {\em ArXiv e-prints}, September 2017.

\bibitem{2018arXiv180505874L}
D.~{Le Quoc}, I.~{Ekin Akkus}, P.~{Bhatotia}, S.~{Blanas}, R.~{Chen},
  C.~{Fetzer}, and T.~{Strufe}.
\newblock {Approximate Distributed Joins in Apache Spark}.
\newblock {\em ArXiv e-prints}, May 2018.

\bibitem{Tachyon}
Haoyuan Li, Ali Ghodsi, Matei Zaharia, Baldeschwieler Eric, Scott Shenker, and
  Ion Stoica.
\newblock Tachyon: Memory throughput i/o for cluster computing frameworks.
\newblock In {\em 7th Workshop on Large-Scale Distributed Systems and
  Middleware}, LADIS'13, 2013.

\bibitem{Tachyon1}
Haoyuan Li, Ali Ghodsi, Matei Zaharia, Scott Shenker, and Ion Stoica.
\newblock Tachyon: Reliable, memory speed storage for cluster computing
  frameworks.
\newblock In {\em Proceedings of the ACM Symposium on Cloud Computing}, SOCC
  '14, pages 6:1--6:15, New York, NY, USA, 2014. ACM.

\bibitem{heterospark_nas15}
Peilong Li, Yan Luo, Ning Zhang, and Yu~Cao.
\newblock Heterospark: A heterogeneous cpu/gpu spark platform for machine
  learning algorithms.
\newblock In {\em Networking, Architecture and Storage (NAS), 2015 IEEE
  International Conference on}, pages 347--348, Aug 2015.

\bibitem{Liu:2017:HCC:3079079.3079089}
Haikun Liu, Yujie Chen, Xiaofei Liao, Hai Jin, Bingsheng He, Long Zheng, and
  Rentong Guo.
\newblock Hardware/software cooperative caching for hybrid dram/nvm memory
  architectures.
\newblock In {\em Proceedings of the International Conference on
  Supercomputing}, ICS '17, pages 26:1--26:10, New York, NY, USA, 2017. ACM.

\bibitem{7980000}
S.~Liu, H.~Wang, and B.~Li.
\newblock Optimizing shuffle in wide-area data analytics.
\newblock In {\em 2017 IEEE 37th International Conference on Distributed
  Computing Systems (ICDCS)}, pages 560--571, June 2017.

\bibitem{Lu}
Xiaoyi Lu, Md. Wasi~Ur Rahman, Nusrat Islam, Dipti Shankar, and Dhabaleswar~K.
  Panda.
\newblock Accelerating spark with rdma for big data processing: Early
  experiences.
\newblock In {\em Proceedings of the 2014 IEEE 22Nd Annual Symposium on
  High-Performance Interconnects}, HOTI '14, pages 9--16, Washington, DC, USA,
  2014. IEEE Computer Society.

\bibitem{maas2016taurus}
Martin Maas, Krste Asanovi{\'c}, Tim Harris, and John Kubiatowicz.
\newblock Taurus: A holistic language runtime system for coordinating
  distributed managed-language applications.
\newblock In {\em Proceedings of the Twenty-First International Conference on
  Architectural Support for Programming Languages and Operating Systems}, pages
  457--471. ACM, 2016.

\bibitem{maas2015trash}
Martin Maas, Tim Harris, Krste Asanovi{\'c}, and John Kubiatowicz.
\newblock Trash day: Coordinating garbage collection in distributed systems.
\newblock In {\em 15th Workshop on Hot Topics in Operating Systems (HotOS XV)},
  2015.

\bibitem{Malewicz:2010:PSL:1807167.1807184}
Grzegorz Malewicz, Matthew~H. Austern, Aart~J.C Bik, James~C. Dehnert, Ilan
  Horn, Naty Leiser, and Grzegorz Czajkowski.
\newblock Pregel: A system for large-scale graph processing.
\newblock In {\em Proceedings of the 2010 ACM SIGMOD International Conference
  on Management of Data}, SIGMOD '10, pages 135--146, New York, NY, USA, 2010.
  ACM.

\bibitem{adobe_icce16}
D.~Manzi and D.~Tompkins.
\newblock Exploring gpu acceleration of apache spark.
\newblock In {\em 2016 IEEE International Conference on Cloud Engineering
  (IC2E)}, pages 222--223, April 2016.

\bibitem{MassieADAM}
Massie, MattNothaft, FrankHartl, ChristopherKozanitis, ChristosSchumacher,
  AndreJoseph, Anthony~D. Patterson, and David~A. Eecs.
\newblock Adam: Genomics formats and processing patterns for cloud scale
  computing.

\bibitem{meng2015mllib}
Xiangrui Meng, Joseph Bradley, Burak Yavuz, Evan Sparks, Shivaram Venkataraman,
  Davies Liu, Jeremy Freeman, DB~Tsai, Manish Amde, Sean Owen, et~al.
\newblock Mllib: Machine learning in apache spark.
\newblock {\em arXiv preprint arXiv:1505.06807}, 2015.

\bibitem{Spark_SQL}
Armbrust Michael, S.~Xin Reynold, Lian Cheng, Huai Yin, Liu Davies, K.~Bradley
  Joseph, Meng Xiangrui, Kaftan Tomer, J.~Franklin Michael, Ghodsi Ali, and
  Zaharia Matei.
\newblock Spark sql: Relational data processing in spark.
\newblock In {\em Proceedings of the 2015 ACM SIGMOD International Conference
  on Management of Data}, SIGMOD '15, Melbourne, Victoria, Australia, 2015.
  ACM.

\bibitem{morales2015samoa}
Gianmarco De~Francisci Morales and Albert Bifet.
\newblock Samoa: Scalable advanced massive online analysis.
\newblock {\em Journal of Machine Learning Research}, 16:149--153, 2015.

\bibitem{moritz2015sparknet}
Philipp Moritz, Robert Nishihara, Ion Stoica, and Michael~I Jordan.
\newblock Sparknet: Training deep networks in spark.
\newblock {\em arXiv preprint arXiv:1511.06051}, 2015.

\bibitem{neumann2011efficiently}
Thomas Neumann.
\newblock Efficiently compiling efficient query plans for modern hardware.
\newblock {\em Proceedings of the VLDB Endowment}, 4(9):539--550, 2011.

\bibitem{Nicolae:2017:LAI:3101627.3101641}
Bogdan Nicolae, Carlos H.~A. Costa, Claudia Misale, Kostas Katrinis, and Yoonho
  Park.
\newblock Leveraging adaptive i/o to optimize collective data shuffling
  patterns for big data analytics.
\newblock {\em IEEE Trans. Parallel Distrib. Syst.}, 28(6):1663--1674, June
  2017.

\bibitem{Nishtala}
Rajesh Nishtala, Hans Fugal, Steven Grimm, Marc Kwiatkowski, Herman Lee,
  Harry~C. Li, Ryan McElroy, Mike Paleczny, Daniel Peek, Paul Saab, David
  Stafford, Tony Tung, and Venkateshwaran Venkataramani.
\newblock Scaling memcache at facebook.
\newblock In {\em Proceedings of the 10th USENIX Conference on Networked
  Systems Design and Implementation}, nsdi'13, pages 385--398, Berkeley, CA,
  USA, 2013. USENIX Association.

\bibitem{Nothaft2015Rethinking}
Frank~Austin Nothaft, Matt Massie, Timothy Danford, Carl Yeksigian, Carl
  Yeksigian, Carl Yeksigian, Jey Kottalam, Arun Ahuja, Jeff Hammerbacher, and
  Michael Linderman.
\newblock Rethinking data-intensive science using scalable analytics systems.
\newblock In {\em ACM SIGMOD International Conference on Management of Data},
  pages 631--646, 2015.

\bibitem{Olston:2008:PLN:1376616.1376726}
Christopher Olston, Benjamin Reed, Utkarsh Srivastava, Ravi Kumar, and Andrew
  Tomkins.
\newblock Pig latin: A not-so-foreign language for data processing.
\newblock In {\em Proceedings of the 2008 ACM SIGMOD International Conference
  on Management of Data}, SIGMOD '08, pages 1099--1110, New York, NY, USA,
  2008. ACM.

\bibitem{Sparrow}
Kay Ousterhout, Patrick Wendell, Matei Zaharia, and Ion Stoica.
\newblock Sparrow: Distributed, low latency scheduling.
\newblock In {\em Proceedings of the Twenty-Fourth ACM Symposium on Operating
  Systems Principles}, SOSP '13, pages 69--84, New York, NY, USA, 2013. ACM.

\bibitem{pu2015low}
Qifan Pu, Ganesh Ananthanarayanan, Peter Bodik, Srikanth Kandula, Aditya
  Akella, Paramvir Bahl, and Ion Stoica.
\newblock Low latency geo-distributed data analytics.
\newblock In {\em Proceedings of the 2015 ACM Conference on Special Interest
  Group on Data Communication}, pages 421--434. ACM, 2015.

\bibitem{Pujol:2010:LES:1851182.1851227}
Josep~M. Pujol, Vijay Erramilli, Georgos Siganos, Xiaoyuan Yang, Nikos
  Laoutaris, Parminder Chhabra, and Pablo Rodriguez.
\newblock The little engine(s) that could: Scaling online social networks.
\newblock In {\em Proceedings of the ACM SIGCOMM 2010 Conference}, SIGCOMM '10,
  pages 375--386, New York, NY, USA, 2010. ACM.

\bibitem{Ramnarayan:2016:SHT:2882903.2899408}
Jags Ramnarayan, Barzan Mozafari, Sumedh Wale, Sudhir Menon, Neeraj Kumar,
  Hemant Bhanawat, Soubhik Chakraborty, Yogesh Mahajan, Rishitesh Mishra, and
  Kishor Bachhav.
\newblock Snappydata: A hybrid transactional analytical store built on spark.
\newblock In {\em Proceedings of the 2016 International Conference on
  Management of Data}, SIGMOD '16, pages 2153--2156, New York, NY, USA, 2016.
  ACM.

\bibitem{10.1007/978-3-319-74690-6_66}
Mostafa~Mohamed Seif, Essam~M. Ramzy~Hamed, and Abd El~Fatah Abdel
  Ghfar~Hegazy.
\newblock Stock market real time recommender model using apache spark
  framework.
\newblock In Aboul~Ella Hassanien, Mohamed~F. Tolba, Mohamed Elhoseny, and
  Mohamed Mostafa, editors, {\em The International Conference on Advanced
  Machine Learning Technologies and Applications (AMLTA2018)}, pages 671--683,
  Cham, 2018. Springer International Publishing.

\bibitem{7930005}
E.~R. Sparks, S.~Venkataraman, T.~Kaftan, M.~J. Franklin, and B.~Recht.
\newblock Keystoneml: Optimizing pipelines for large-scale advanced analytics.
\newblock In {\em 2017 IEEE 33rd International Conference on Data Engineering
  (ICDE)}, pages 535--546, April 2017.

\bibitem{C_store}
Mike Stonebraker, Daniel~J. Abadi, Adam Batkin, Xuedong Chen, Mitch Cherniack,
  Miguel Ferreira, Edmond Lau, Amerson Lin, Sam Madden, Elizabeth O'Neil, Pat
  O'Neil, Alex Rasin, Nga Tran, and Stan Zdonik.
\newblock C-store: A column-oriented dbms.
\newblock In {\em Proceedings of the 31st International Conference on Very
  Large Data Bases}, VLDB '05, pages 553--564. VLDB Endowment, 2005.

\bibitem{talwalkar2012mlbase}
A~Talwalkar, T~Kraska, R~Griffith, J~Duchi, J~Gonzalez, D~Britz, X~Pan,
  V~Smith, E~Sparks, A~Wibisono, et~al.
\newblock Mlbase: A distributed machine learning wrapper.
\newblock In {\em NIPS Big Learning Workshop}, 2012.

\bibitem{Tang:2012:EEP:2197076.2197192}
Shanjiang Tang, Ce~Yu, Jizhou Sun, Bu-Sung Lee, Tao Zhang, Zhen Xu, and Huabei
  Wu.
\newblock Easypdp: An efficient parallel dynamic programming runtime system for
  computational biology.
\newblock {\em IEEE Trans. Parallel Distrib. Syst.}, 23(5):862--872, May 2012.

\bibitem{Hive}
A.~Thusoo, J.S. Sarma, N.~Jain, Zheng Shao, P.~Chakka, Ning Zhang, S.~Antony,
  Hao Liu, and R.~Murthy.
\newblock Hive - a petabyte scale data warehouse using hadoop.
\newblock In {\em Data Engineering (ICDE), 2010 IEEE 26th International
  Conference on}, pages 996--1005, March 2010.

\bibitem{7912039}
Rajeshwari U and B.~S. Babu.
\newblock Real-time credit card fraud detection using streaming analytics.
\newblock In {\em 2016 2nd International Conference on Applied and Theoretical
  Computing and Communication Technology (iCATccT)}, pages 439--444, July 2016.

\bibitem{YARN}
Vinod~Kumar Vavilapalli, Arun~C. Murthy, Chris Douglas, Sharad Agarwal, Mahadev
  Konar, Robert Evans, Thomas Graves, Jason Lowe, Hitesh Shah, Siddharth Seth,
  Bikas Saha, Carlo Curino, Owen O'Malley, Sanjay Radia, Benjamin Reed, and
  Eric Baldeschwieler.
\newblock Apache hadoop yarn: Yet another resource negotiator.
\newblock In {\em Proceedings of the 4th Annual Symposium on Cloud Computing},
  SOCC '13, pages 5:1--5:16, New York, NY, USA, 2013. ACM.

\bibitem{Shivaram}
Shivaram Venkataraman, Aurojit Panda, Ganesh Ananthanarayanan, Michael~J.
  Franklin, and Ion Stoica.
\newblock The power of choice in data-aware cluster scheduling.
\newblock In {\em Proceedings of the 11th USENIX Conference on Operating
  Systems Design and Implementation}, OSDI'14, pages 301--316, Berkeley, CA,
  USA, 2014. USENIX Association.

\bibitem{venkataramansparkr}
Shivaram Venkataraman, Zongheng Yang, Eric~Liang Davies~Liu, Hossein Falaki,
  Xiangrui Meng, Reynold Xin, Ali Ghodsi, Michael Franklin, Ion Stoica, and
  Matei Zaharia.
\newblock Sparkr: Scaling r programs with spark.

\bibitem{Weil:2006:CSH:1298455.1298485}
Sage~A. Weil, Scott~A. Brandt, Ethan~L. Miller, Darrell D.~E. Long, and Carlos
  Maltzahn.
\newblock Ceph: A scalable, high-performance distributed file system.
\newblock In {\em Proceedings of the 7th Symposium on Operating Systems Design
  and Implementation}, OSDI '06, pages 307--320, Berkeley, CA, USA, 2006.
  USENIX Association.

\bibitem{wiewiorka2014sparkseq}
Marek~S Wiewi{\'o}rka, Antonio Messina, Alicja Pacholewska, Sergio Maffioletti,
  Piotr Gawrysiak, and Micha{\l}~J Okoniewski.
\newblock Sparkseq: fast, scalable and cloud-ready tool for the interactive
  genomic data analysis with nucleotide precision.
\newblock {\em Bioinformatics}, 30(18):2652--2653, 2014.

\bibitem{Shark}
Reynold~S. Xin, Josh Rosen, Matei Zaharia, Michael~J. Franklin, Scott Shenker,
  and Ion Stoica.
\newblock Shark: Sql and rich analytics at scale.
\newblock In {\em Proceedings of the 2013 ACM SIGMOD International Conference
  on Management of Data}, SIGMOD '13, pages 13--24, New York, NY, USA, 2013.
  ACM.

\bibitem{Yan:2016:TTC:2987550.2987576}
Ying Yan, Yanjie Gao, Yang Chen, Zhongxin Guo, Bole Chen, and Thomas
  Moscibroda.
\newblock Tr-spark: Transient computing for big data analytics.
\newblock In {\em Proceedings of the Seventh ACM Symposium on Cloud Computing},
  SoCC '16, pages 484--496, New York, NY, USA, 2016. ACM.

\bibitem{RDD}
Matei Zaharia, Mosharaf Chowdhury, Tathagata Das, Ankur Dave, Justin Ma, Murphy
  McCauley, Michael~J. Franklin, Scott Shenker, and Ion Stoica.
\newblock Resilient distributed datasets: A fault-tolerant abstraction for
  in-memory cluster computing.
\newblock In {\em Proceedings of the 9th USENIX Conference on Networked Systems
  Design and Implementation}, NSDI'12, pages 2--2, Berkeley, CA, USA, 2012.
  USENIX Association.

\bibitem{Spark}
Matei Zaharia, Mosharaf Chowdhury, Michael~J. Franklin, Scott Shenker, and Ion
  Stoica.
\newblock Spark: Cluster computing with working sets.
\newblock In {\em Proceedings of the 2Nd USENIX Conference on Hot Topics in
  Cloud Computing}, HotCloud'10, pages 10--10, Berkeley, CA, USA, 2010. USENIX
  Association.

\bibitem{Discretized_Streams}
Matei Zaharia, Tathagata Das, Haoyuan Li, Timothy Hunter, Scott Shenker, and
  Ion Stoica.
\newblock Discretized streams: Fault-tolerant streaming computation at scale.
\newblock In {\em Proceedings of the Twenty-Fourth ACM Symposium on Operating
  Systems Principles}, SOSP '13, pages 423--438, New York, NY, USA, 2013. ACM.

\bibitem{Zhang}
Hao Zhang, Bogdan~Marius Tudor, Gang Chen, and Beng~Chin Ooi.
\newblock Efficient in-memory data management: An analysis.
\newblock {\em Proc. VLDB Endow.}, 7(10):833--836, June 2014.

\bibitem{Zhang:2018:ROS:3190508.3190534}
Haoyu Zhang, Brian Cho, Ergin Seyfe, Avery Ching, and Michael~J. Freedman.
\newblock Riffle: Optimized shuffle service for large-scale data analytics.
\newblock In {\em Proceedings of the Thirteenth EuroSys Conference}, EuroSys
  '18, pages 43:1--43:15, New York, NY, USA, 2018. ACM.

\bibitem{Splash}
Yuchen Zhang and Michael~I. Jordan.
\newblock Splash: User-friendly programming interface for parallelizing
  stochastic algorithms.
\newblock {\em CoRR}, abs/1506.07552, 2015.

\bibitem{zhang2016kira}
Zhao Zhang, Kyle Barbary, Frank~A Nothaft, Evan~R Sparks, Oliver Zahn,
  Michael~J Franklin, David~A Patterson, and Saul Perlmutter.
\newblock Kira: Processing astronomy imagery using big data technology.
\newblock {\em IEEE Transactions on Big Data}, 2016.

\bibitem{zhang2015scientific}
Zhao Zhang, Kyle Barbary, Frank~Austin Nothaft, Evan Sparks, Oliver Zahn,
  Michael~J Franklin, David~A Patterson, and Saul Perlmutter.
\newblock Scientific computing meets big data technology: An astronomy use
  case.
\newblock In {\em Big Data (Big Data), 2015 IEEE International Conference on},
  pages 918--927. IEEE, 2015.

\end{thebibliography}

\end{document}